\documentclass[11pt,a4paper]{article}

\usepackage{jheppub}
\usepackage{epsfig,amssymb,amsmath,psfrag}

\usepackage[T1]{fontenc}
\usepackage{psfrag}

\DeclareMathOperator{\sech}{sech}

\catcode`\@=11
\def \be  {\begin{equation}}
\def \ee  {\end{equation}}
\def \ba  {\begin{eqnarray}}
\def \ea  {\end{eqnarray}}
\def \baa {\begin{eqnarray*}}
\def \eaa {\end{eqnarray*}}
\def \bb  {\begin {thebibliography} }
\def \eb  {\end{thebibliography}}

\def \Im {\mathop{\rm Im}\nolimits}
\def \Re {\mathop{\rm Re}\nolimits}
\def \e  {\mathop{\rm e}\nolimits}
\def\phione{\phi}

\newcommand{\ft}[2]{{\textstyle\frac{#1}{#2}}}

\begin{document}

\bibliographystyle{JHEP}


\title{Excited Hexagon Wilson Loops for \\[3mm]  Strongly Coupled ${\cal N}=4$ SYM}

\author[a]{J.~Bartels}
\author[a]{J.~Kotanski}
\author[b,c]{V.~Schomerus}

\emailAdd{joachim.bartels@desy.de}
\emailAdd{jan.kotanski@desy.de}
\emailAdd{volker.shomerus@desy.de}

\affiliation[a]{II. Institute Theoretical Physics,\\
Hamburg University, Germany}
\affiliation[b]{DESY Theory Group,\\
Hamburg, Germany}
\affiliation[c]{Department of Physics and Astronomy,\\
 UBC, Vancouver, Canada}

\abstract{This work is devoted to the six-gluon scattering amplitude in
strongly coupled ${\cal N}=4$ supersymmetric Yang-Mills theory. At weak
coupling, an appropriate high energy limit of the so-called
remainder function, i.e. of the deviation from the BDS formula,
may be understood in terms of the lowest eigenvalue of the BFKL
hamiltonian. According to Alday et al., amplitudes in the strongly
coupled theory can be constructed through an auxiliary
1-dimensional quantum system. We argue that certain excitations of
this quantum system determine the Regge limit of the remainder
function at strong coupling and we compute its precise value.}

\keywords{Keywords: AdS/CFT, gluon amplitudes, Regge limit, thermodynamic Bethe Ansatz}

\arxivnumber{1009.3938}

\maketitle

\section{Introduction}

The correspondence between ${\cal N}=4$ supersymmetric Yang-Mills theory
and string theory on an $AdS_5 \times S_5$ geometry is the most
studied example of a gauge/string duality
\cite{Polyakov:1980ca,Maldacena:1997re,Witten:1998qj,Gubser:1998bc}.
After the impressive progress that has been made in computing
anomalous dimensions of gauge invariant operators, see e.g.
\cite{Arutyunov:2009zu,Gromov:2009tv,Bombardelli:2009ns,
Gromov:2009bc,Arutyunov:2009ur,Hegedus:2009ky,Gromov:2009zb,
Arutyunov:2009ux,
Gromov:2009tq}
and references
therein, the focus has recently shifted to the analysis of planar
scattering amplitudes.

On the gauge theory side, the discussion was stimulated by the
intriguing BDS formula of Bern, Dixon and Smirnov \cite{Bern:2005iz}. It
encapsulates the known infrared and collinear behavior of
$n$-particle maximally helicity violating (MHV) amplitudes in the
planar approximation. The authors of \cite{Bern:2005iz} conjectured the
BDS formula to determine the amplitudes at each loop order $L\ge2$
up to some additive finite function $R^{(n)}$ of the kinematic
variables. $R^{(n)}$ has been dubbed the 'remainder function'.
Both string and gauge theory arguments suggest that the remainder
function $R^{(n)}$ vanishes for $n=4,5$. In the weakly coupled theory,
perturbative computations have uncovered an important new symmetry
of scattering amplitudes \cite{Drummond:2007au}. Assuming this
so-called {\em dual conformal symmetry}, the remainder functions
$R^{(n)}$ can only depend on conformal cross ratios, i.e.\ on
conformal invariant combinations of the usual kinematic variables.
Since there are no such cross ratios for $n=4,5$, the
corresponding remainder functions have to be trivial. In other
words, dual conformal invariance predicts that the BDS formula is
exact for $n=4,5$ to all loop orders. This prediction was
confirmed by a string theory computation of the leading term at
strong coupling. We shall say a bit more about the string
theoretic analysis below.

On the other hand, the remainder function $R^{(n)}$ is known to be
non-zero for $n>5$ and beyond one loop
\cite{Drummond:2007bm,Bern:2008ap}. A first explicit disagreement
between the BDS formula and calculations of Yang-Mills scattering
amplitudes in the leading logarithmic high energy approximation
was pointed out in \cite{Bartels:2008ce,Bartels:2008sc}. This
outcome has been confirmed by a numerical analysis of existing
two-loop calculations for the six-point function
\cite{Schabinger:2009bb}.

The analysis in \cite{Bartels:2008ce,Bartels:2008sc} illustrates
the importance of the high energy limit (Regge limit) which, in
particular, exhibits much of the subtle analytic structure of
multiparticle amplitudes. Let us briefly recapitulate the main
results. There are three different kinematical regions which are
of interest. In the usual `physical region' all energies of the
process are positive. The opposite case in which all energies are
negative is referred to as the `Euclidean region'. Most of the
analysis we shall discuss in this work, however, deals with a
third region that contains both positive and negative energies,
see eq.\ \eqref{newregime} and Fig. 1 for a precise definition of
this `mixed region'. It is the mixed region that offers the
clearest view on the discrepancy between the BDS formula and the
existing high energy calculations of gluon amplitudes. In fact,
the six-gluon scattering amplitude in the mixed region includes a
Regge-cut contribution which cannot be reproduced from the BDS
formula. For the physical regime, the failure of the BDS formula
manifests itself in a violation of the Steinmann relation. No
disagreement is visible in the Euclidean region.

These findings imply that the remainder function $R^{(6)}$ vanishes in
the high energy limit of both the Euclidean and the physical
region. Upon analytic continuation into the mixed region, however,
the high energy limit should be nonzero and, in the leading
logarithmic approximation, must coincide with the contribution
given in \cite{Bartels:2008sc}. This emphasizes the importance of the
correct analytic structure. In a recent study \cite{Lipatov:2010qg}, a
convenient analytic expression \cite{Goncharov:2010jf} for the exact two-loop calculation
of the six-point function
\cite{DelDuca:2009au,DelDuca:2010zg} was used to
perform the relevant analytic continuation. The results are in
full agreement with \cite{Bartels:2008sc}. This settles the remainder
function in the two-loop approximation, and it supports the
all-order leading log generalization in \cite{Bartels:2008ce,Bartels:2008sc}.

The leading logarithmic approximation in the mixed
region exhibits an important feature. As a function of the energy
variable $s_2$ (see Fig.1) the six-gluon amplitude Amp$^\prime$
\footnote{The $\prime$ is placed on this amplitude to remind us
the high energy limit is taken in the mixed rather than the
physical region.} contains a Regge cut term with a power-like behavior
\ba \mbox{Amp}^{\prime} \ \sim\  s_2^{- E_2}\,, \ea
where the (negative-valued) exponent $E_2$ is the lowest
eigenvalue of the BFKL color-octet Hamiltonian $H_2$ for a
two-gluon system. This term is missing in the BDS-amplitude and
therefore should be connected with the remainder function $R^{(6)}$.
The BFKL color-octet Hamiltonian coincides with the Hamiltonian
of an integrable open spin chain \cite{Lipatov:2009nt}. Similar
results hold for scattering amplitudes involving $n>6$ gluons. In
a processes $2 \to 2n$ there exist Regge cut contributions with up
to $n$ gluons and the corresponding spin chains consist of $n$
spins. Hence, the weak coupled theory provides direct evidence
for integrability in the high energy behavior of planar scattering
amplitudes.

Having reviewed all these results from gauge theory it is natural
to ask what string theory has to say about the high energy limit
of the remainder function $R^{(6)}$. In order to understand how the
issue can be addressed, we need to briefly sketch the development
that was initiated by the work \cite{Alday:2007hr} of Alday and
Maldacena. The main insight of this paper was to identify the
leading strong coupling contribution to an n-gluon amplitude with
the area $A_n$ of some 2-dimensional surface $S_n$ inside $AdS_5$.
According to the prescription of \cite{Alday:2007hr}, $S_n$ ends
on a piecewise light-like polygon in the boundary of $AdS_5$. The
light-like segments of this polygon are given by momenta $p_j$ of
the external gluons. For $n=4$ it is possible to find the surface
explicitly and the resulting amplitude matches the prediction from
the BDS formula. Constructing $S_n$ for $n> 5$, however, turned
out to be a rather difficult problem, at least for finite $n$ and
generic choice of the external momenta. The issue was resolved
through a series of papers
\cite{Alday:2009yn,Alday:2009dv,Alday:2010vh} in which the area of
$S_n$ is related to the free energy of some auxiliary quantum
integrable system. More precisely, it was argued that $A_n$ may be
computed from a family of functions $Y_{s,j}$ with $s = 1,2,3$ and
$j = 1, \dots ,n-5$. The latter can be determined by solving a set
of coupled non-linear integral equations. Very similar
mathematical structures are actually familiar from the study of
{\em ground states} in 1-dimensional quantum integrable systems.
Moreover, the functional $A_n$ resembles expressions for the free
energy of such systems.

We employ this auxiliary quantum integrable system to study the
Regge limit of the remainder function $R^{(6)}$ at strong
coupling. The non-linear integral equations for the $Y$-functions
we have mentioned in the previous paragraph provide an explicit
set of equations which enable us to take the high energy limit and
to investigate the analytic continuation between the physical and
the mixed region. As the main result, we determine the value of
the exponent $E_2$ at strong coupling,
 \be E_2 \ = \
\frac{\sqrt{\lambda}}{2\pi} \left( \sqrt 2 + \frac12 \log ( 3 + 2
\sqrt{2}) \right) \ \ . \ee
In passing to the mixed regime, we need to analytically continue
the parameters of the auxiliary quantum integrable system. This
continuation leads to a new system of equations with additional
terms. The structure of the new integral equations and the
corresponding modifications of the relevant energy functional are
again familiar from the theory of quantum integrable systems. In
that context they are used to describe {\em excitations} of the
system. Our formula for the value of $E_2$ at strong coupling
assumes that we can perform the analytic continuation into the
mixed regime within the parameter-space of the quantum integrable
system and that the corresponding amplitude is given by the
exponential of the energy functional, as in the Euclidean regime.

This paper is organized as follows. In section 2 we define the
kinematic regions and the path of analytic continuation for the
kinematic variables. Section 3 starts with a brief summary of
relevant results from the work of Alday et. al. As a first
exercise we then verify that the Regge limit of the remainder
function $R^{(6)}$ vanishes in the physical regime (up to an
irrelevant constant). Sections 4 and 5 contain the main part of
our analysis. In these two sections we perform the analytic
continuation, discuss the origin of the additional terms and
calculate their contribution to the amplitude. For pedagogical
reasons, we first illustrate the general arguments with a simpler
example that is taken from the work \cite{Dorey:1996re,Dorey:1997rb} of Dorey et al.
The generalization to the system of Alday et al. is
straightforward. Section 5 is devoted to the analysis of the
resulting equations. With strong support from numerics we argue
that the transition to the mixed regime is associated with the
simplest kind of excitations in the auxiliary integrable system.
Once this is established, the precise value of $E_2$ can be
calculated analytically. A few technical details are contained in
two appendices.

\section{Multi-Regge kinematics of six-gluon amplitudes}

We consider the scattering of six gluons with momenta $p_j, j = 1,
\dots, 6$ satisfying the on-shell condition $p_j^2 = 0$. From
these momenta we can form Lorentz invariant combinations of the
form
\be x^2_{ij}\  = \ (p_{i+1} + \dots + p_j)^2 \ .
\label{distances}
\ee
Throughout this note we shall use the metric $\eta =${\it
diag\/}$(-,+,+,+)$. Furthermore, we extend the range of the index
$j$ on $p_j$ to the integers with the periodicity condition $p_j
\equiv p_{j+6}$. Due to momentum conservation the quantities
$x^2_{ij}$ are symmetric in their indices, i.e.\ the obey
$x^2_{ij} = x_{ji}^2$. Six-point amplitudes of a 4-dimensional
gauge theory depend on eight kinematical invariants. Hence, there
exist many relations between the $x^2_{ij}$ we introduced above.

\begin{figure}
\begin{center}
{
}
{\epsfysize6.5cm \epsfbox{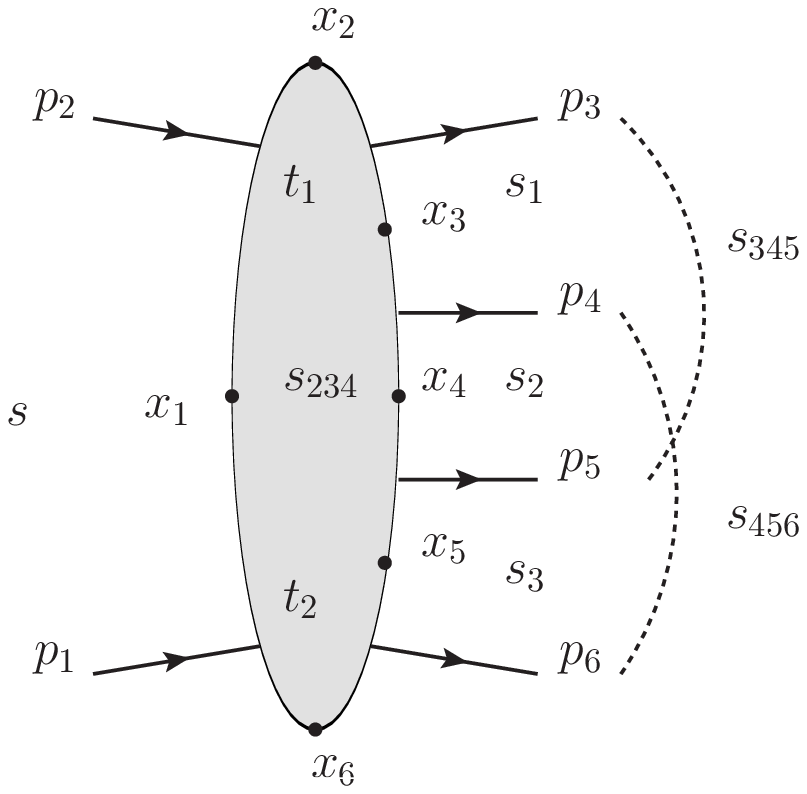}
\;\epsfysize6.5cm \epsfbox{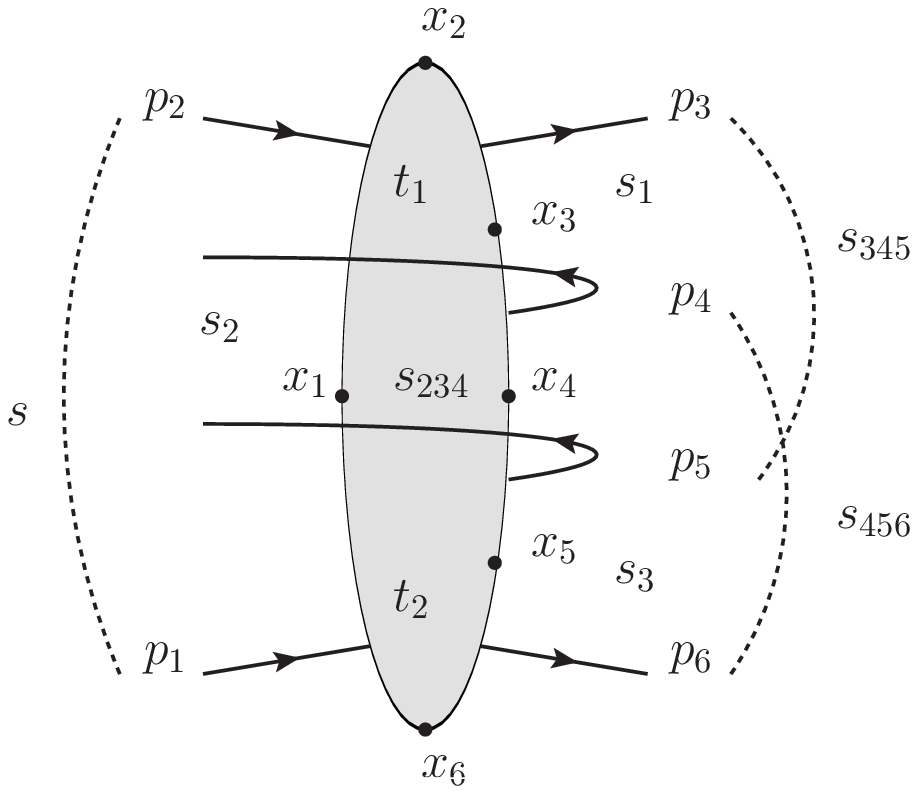}}
\end{center}
\caption{Kinematic configuration for the multi-Regge limit before (on the left) and after (on the right) the analytical continuation with the kinematic invariants from eq.~\eqref{stinv}. Momenta of are denoted by $p_i$ while dual coordinates of the Wilson loop by $x_i$.}
\label{fig:mrl}
\end{figure}

In the literature, six-point amplitudes are usually parametrized
in terms of the following nine invariants, as shown schematically
in figure \ref{fig:mrl},
\ba
s\ =\ - x_{26}^2\,,
\quad  \quad s_1\ =\ - x_{24}^2\,, & &  s_2
\ =\ - x_{35}^2\,,\quad \quad s_3\ = \ - x_{46}^2\,, \nonumber \\[2mm]
 t_1\ =\ - x_{13}^2\,, & &  t_2\ =\ - x_{15}^2\,,
\label{stinv} \\[2mm]
s_{345}\ =\ - x_{25}^2\,, \quad \quad s_{234}\!&=&\! - x_{14}^2\,, \quad
\quad s_{456}\ =\ - x_{36}^2\nonumber \ . \ea
Variables denoted by $s$ or $t$ are referred to as energies
and momentum transfers, resp. In the Regge limit, energies are large
(and can be positive or negative), whereas momentum transfer variables
are negative and finite.
As an example, a positive energy $s$ implies
the negative distance $x_{26}^2$.
Since we have only eight independent variables, one of the above
can be eliminated with the help of the Gram determinant, namely
$\det([p_i \cdot p_j]_{i,j=1,\ldots,5})=0$. Since each of the
products $[p_i \cdot p_j]$ may be expressed through the
kinematical invariants defined in eqs.\ \eqref{stinv}, we can use
the Gram determinant relation to eliminate one of these nine
invariants.

In the physical regime, all the `s-like' kinematical invariants
are positive while the `t-like' variables are negative, i.e.\
\be \label{physical} s,s_i,s_{456},s_{345}\ > \ 0 \quad \mbox{ and
} \quad
   t_i,s_{234} < 0 \ \ . \ee

Finite quantities in a theory with dual conformal invariance can
only depend on cross-ratios. For a six-point amplitude, there are
three independent conformal cross-ratios which are defined as
\be u_1\ =\ \frac{x_{13}^2 x_{46}^2}{x_{14}^2 x_{36}^2}\ =\
\frac{t_1 s_3}{s_{234} s_{456}}\,, \quad u_2\ =\ \frac{x_{24}^2
x_{51}^2}{x_{25}^2 x_{41}^2}\ = \ \frac{s_1 t_2}{s_{345}
s_{234}}\,, \quad u_3\ =\ \frac{x_{35}^2 x_{62}^2}{x_{36}^2
x_{52}^2}\ =\ \frac{s_2 s }{s_{456} s_{345}}\, .
\label{eq:crossratio}
\ee
For later use we have expressed them both in terms of the general
kinematical invariants $x_{ij}$ and in terms of the nine special
invariants that were defined in eq.\ \eqref{stinv}.

In the next section we shall look at the multi-Regge limit for
six-point functions which, by definition, explores the regime
\ba & & s\ \gg\ s_1 , s_2 , s_3\ \gg\ -t_1 , -t_2 , -s_{234} \
\sim\ \mbox{fixed}\ .  \label{Reggenormal} \ea
The conformal cross-ratios are functions of the kinematical
invariants and one can study there behavior in the Regge regime.
It turns out that
\ba u_1 \ \sim \ u_2 \ \sim \ 1-u_3\  \to\  0 \label{Reggenormalu}
\ea
vanish in the same order (see Appendix \ref{ap:rl}). Hence, the
leading asymptotics of any conformal invariant of the momenta
$p_i$ may be parametrized by the two quantities
\ba
 \tilde u_1 \ = \ u_1/(1-u_3) \quad \mbox{ and } \quad
   \tilde u_2 \ = \ u_2/(1-u_3) \ \ .
\label{utilde}
\ea
In this simple multi-Regge limit, the six gluon scattering
amplitude actually agrees with the answer that is predicted by the
BDS formula. This is because the limiting values
$(u_1,u_2,u_3)_{\textrm{Regge}} = (0,0,1)$, corresponds to a
special case of the collinear limit, namely $u_1=0$, $u_3=1-u_2$
where $u_2 \to 0$. In the collinear limit, the six-point amplitude
reduces a 5-point amplitude which is known to be BDS exact. Hence,
the Regge limit \eqref{Reggenormal} of the BDS formula for six
gluons needs no corrections and therefore is of no further
interest to us, at least for the moment.

As we have outlined in the introduction, however, there exists
another Regge limit in which the BDS formula requires corrections,
i.e.\ the remainder function does not vanish. Reaching this
limiting regime requires to pass to a  new kinematic region which
is defined as
\be \label{newregime}
   s,s_2\, > \ 0 \quad \mbox{ and } \quad
 s_1,s_3, s_{345} s_{456}\, < 0\,,\ t_1,t_2,s_{234}\, < 0  \ \ .
\ee
In the corresponding multi-Regge limit all energy variables are
much larger than the momentum transfer variables,
\ba & & s\ \gg\ s_2 , -s_1, -s_3\ \gg\ -t_1 , -t_2 , -s_{234} \
\sim\ \mbox{fixed}\,. \label{Reggefunny} \ea

In order to reach this regime, one either starts from the region
where all invariants are negative and then continues in $s$ and
$s_2$ or, alternatively, one first goes to the physical region and
then continues all the following energy variables analytically
along the following simple path
\be
\label{paths}
s_1(\varphi) \ =\ e^{i\varphi} s_1
\,, \quad
s_3(\varphi) \ =\ e^{i\varphi} s_3
\quad \mbox{ and } \quad
s_{456}(\varphi) \ = \ e^{i \varphi} s_{456}
\,, \quad
s_{345}(\varphi) \ = \ e^{i \varphi} s_{345}\,,
\ee
where $\varphi \in [0,\pi]$ and all the other kinematical
invariants are left unaltered. In the process of the continuation,
the conformal cross ratios $u_1$ and $u_2$ are left invariant
while $u_3$ behaves as
\be  {\cal C}_u \ : \quad
 u_3(\varphi) \ = \ e^{-2i \varphi} u_3\ .
\label{pathu}
\ee
At the final point $\varphi = \pi$ all three cross-ratios come
back to their original values. Along the way, the cross-ratio
$u_3$ moves along a full circle.

Our analysis below starts with some conformal amplitude $A(u_i)$
in the physical regime. Its precise form will be described in the
next section. In a first step we want to continue this amplitude
analytically along the path \eqref{pathu}. Since $A(u_i)$ is not
single valued in the complex $u-$space, the resulting amplitude
$A^{\cal C}(u_i)$ differs from the amplitude we started with. Then
we perform the Regge limit. This leaves us with a function
$a(\tilde u_1,\tilde u_2)$ that describes the leading Regge
asymptotics of $A^{\cal C}$. It is this quantity we are actually
after.

The amplitude we shall subject to this analysis is taken from the
recent work of Alday et al. By construction, it is initially
defined in the Euclidean region in which all the kinematical
invariants \eqref{stinv} are negative. In order to go to the
physical region \eqref{physical} one needs to continue
$s,s_i,s_{345}$ and $s_{456}$ analytically while keeping $t_i$ and
$s_{234}$ fixed. This continuation may be performed such that all
three conformal cross-ratios $u_i$ remain fixed. Hence, conformal
amplitudes do not require analytical continuation in order to pass
from the Euclidean to the physical region.

\section{Six gluon amplitudes at strong coupling}

In the first part of this subsection, we shall briefly recall the
prescription of Alday et al. for six gluon amplitudes at strong 't
Hooft  coupling. The second subsection contains an analysis of the
Regge limit in the physical regime. We shall demonstrate that the
Regge limit of the six gluon amplitude is BDS exact, in agreement
with general arguments.

\subsection{Review of results from Alday et al.}
According to the work of Alday et al. \cite{Alday:2009dv}, the strong coupling
limit of a six gluon amplitude in ${\cal N}=4$ supersymmetric
Yang-Mills theory is obtained through exponentiation of the
quantity
\be A^{(6)}\ =\ A^{(6)}_{\textrm{div}} +
A^{(6)}_{\textrm{BDS-like}} + A^{(6)}_{\textrm{free}} +
A^{(6)}_{\textrm{period}}
 \ ,
\ee at least up to a constant that is independent of the
kinematics. The various pieces on the right hand side are all
known explicitly. In particular the divergent contribution is
given by
\be A^{(6)}_{\textrm{div}}\ =\ \frac1{8} \sum_i \log^2(\epsilon^2
x^2_{i,i+2})\ , \ee
where $\epsilon$ is the cut-off. The BDS-like amplitude is also
relatively simple. In the case of six gluon scattering, it is
given by
\be A^{(6)}_{\textrm{BDS-like}}\ =\ -\frac1{8} \sum_{i=1}^6 \left(
\log^2 x^2_{i,i+2}+ \sum_{k=0}^{2} (-1)^{k+1} \log x^2_{i,i+2}
\log x^2_{i+2k+1,i+2k+3} \right) \ . \ee
For us, it is most relevant to have an expression for the
difference between the usual BDS and the BDS-like amplitudes. This
difference is given by
\be \Delta^{(6)} \ = \
A^{(6)}_{\textrm{BDS}}-A^{(6)}_{\textrm{BDS-like}}\ =\
\sum_{i=1}^3 \left( \frac1{8}\log^2 u_i+\frac1{4}
\text{Li}_2(1-u_i) \right)\ .
\label{eq:Delta}
\ee
The conformal cross-ratios $u_i$ that appear in this expression
were defined in the previous section.

The two remaining contributions to the total amplitude cannot be
written directly in terms of the three cross-ratios $u_i$.
Instead, they involve a new set of parameters $m,C,\phi$ that are
related to the $u_i$ through some functional relations to be
spelled out below. Given $m,C,\phi$ Alday et al. instruct us to
solve the following system of non-linear integral equations for
three functions $Y_i = Y_i(\theta), i =1,2,3$,
\ba \log Y_{2}(\theta) & =& -m \sqrt{2} \cosh (\theta-i \phione) -
2 \int_{-\infty} ^{\infty} d \theta'  K_1(\theta-\theta' ) \log
(1+ Y_{2}(\theta')) \nonumber \\ && - \int_{-\infty} ^{\infty} d
\theta' K_2(\theta-\theta') \log ((1+ Y_{1}(\theta') )(1+
Y_{3}(\theta')))\,, \label{NLIE1}
\ea
\ba \log  Y_{2\pm1}(\theta)&=& -m\cosh(\theta-i \phione) \pm C -
\int_{-\infty} ^{\infty} d \theta'  K_2(\theta-\theta' ) \log (1+
Y_{2}(\theta')) \nonumber
\\ && - \int_{-\infty} ^{\infty} d \theta' K_1(\theta-\theta')
\log ((1+ Y_{1}(\theta'))(1+ Y_{3}(\theta')) ) \ .
\label{NLIE2}
\ea
These formulas can be used as long as $|\Im(\theta-i \phione)|<
\pi/2$. For values of $\theta$ outside this strip, one can employ
certain recurrence relation to determine $Y_i$; details are given in the
Appendix \ref{ap:he}.
The equations \eqref{NLIE1} and \eqref{NLIE2} involve two integral
kernels $K_1$ and $K_2$ that
are given by
\be K_1(\theta)\ =\ \frac{1}{2 \pi \cosh \theta}\ ,\quad
K_2(\theta)\ =\ \frac{\sqrt{2} \cosh \theta}{ \pi \cosh 2 \theta}
\,. \label{Ki} \ee

Suppose for the moment that we can solve these equations somehow.
The resulting functions $Y_i(\theta)$ clearly depend on the
parameters $m,C$ and $\phi$ in some complicated way. It is now
possible to express $m,C,\phi$ through the cross ratios $u_i$ by
solving the following three equations
\ba
\label{uYrelation}
u_1&=&\frac{x^2_{13}x^2_{46}}{x^2_{14}x^2_{36}}\ = \
\left(1+\frac1{Y_{2}(\theta=-i\pi/4)}\right)^{-1}\,,\nonumber \\[2mm]
u_2&=&\frac{x^2_{24}x^2_{15}}{x^2_{25}x^2_{14}}\ =\
\left(1+\frac1{Y_{2}(\theta=i\pi/4)}\right)^{-1}\,,\label{ufrompar} \\[2mm]
u_3&=&\frac{x^2_{35}x^2_{26}}{x^2_{36}x^2_{25}}\ =\
\left(1+\frac1{Y_{2}(\theta=-3i\pi/4)}\right)^{-1}\,. \nonumber \ea
The last equation contains the function $Y_2$ outside the strip
$|Im(\theta-i \phione)|<\pi/2$. With the recurrence relations
described in the Appendix \ref{ap:he} we rewrite this equation in
the following form
 \be u_3\ =\ \left. \frac{ (1-u_2)u_2
}{(1+Y_{1})u_2 +(u_2+Y_{1}) Y_{3}}\right|_{\theta =0} \ =\ \left.
\frac{ (1-u_1)u_1 }{(1+Y_{1})u_1 +(u_1+Y_{1}) Y_{3}}
\right|_{\theta = 0}\,.
 \label{uYrelations2} \ee
Once $m = m(u_i), C = C(u_i)$ and $\phi = \phi(u_i)$ are known, we
can construct the two missing terms of the total amplitude. The
so-called free energy is given by
\be A^{(6)}_{\textrm{free}} \ = \int \frac{d \theta}{2 \pi} m
\cosh \theta \log \left[ (1+Y_{1}(\theta+ i \phione))
(1+Y_{3}(\theta+ i \phione)) (1+Y_{2}(\theta+ i
\phione))^{\sqrt{2}} \right]\ . \label{Afree}  \ee
The final contribution to the amplitude depends on the cross
ratios only through the parameter $m = m(u_i)$,
\be A^{(6)}_{\textrm{period}} \ = \ \frac{1}{4} \, m^2 \ . \ee
This concludes our description of the input from the work of Alday
et al.

\subsection{Expansion for large masses}

It is instructive to first study the Regge limit in the physical
region and to verify that the remainder function does indeed
vanish (up to a constant). Most of the formulas in this subsection
have appeared in the literature before, see
\cite{Alday:2010ku,Hatsuda:2010vr}. We repeat them here mostly
for pedagogical reasons.

In the region \eqref{physical} one can show that the multi-Regge
limit is mapped onto large values of
$m$, while $C$ remains constant and $\phi$ goes to zero. To see
this we start from eqs.\ \eqref{uYrelation}, \eqref{uYrelations2}
and observe that our Regge limit \eqref{Reggenormalu} forces the
values $Y_2(\pm i \pi/4)$, $Y_1(0)$, $Y_3(0)$ to approach zero at
the same rate, i.e.\ the right hand sides of eqs.\ \eqref{NLIE1}
and \eqref{NLIE2} must become large and negative. As long as
$|\phi|<\pi/4$, the parameter $m$ is forced to become large and
the first terms on the right hand side dominate, i.e.\
\ba \log Y_{2}(\theta) & \approx& -m \sqrt{2} \cosh (\theta-i
\phione)\,, \label{largemYtwo}  \\[2mm]
\log Y_{2\pm1}(\theta)&\approx& -m\cosh(\theta-i \phione) \pm C \,
. \label{largemYone} \ea
As we shall demonstrate further below, these approximations are
justified since the integrals on the right hand side of eqs.\
\eqref{NLIE1} and \eqref{NLIE2} remain finite. This will no longer
be that case after analytic continuation into the kinematic region
\eqref{newregime}. Taking the difference of the  two equations
\eqref{NLIE2} we obtain
\ba Y_3(0)/Y_1(0) \ =\  e^{2C}\ . \ea
As we argued above, the Regge limit requires both $Y_1(0)$ and
$Y_3(0)$ to approach zero at the same rate so that their ratio
stays fixed. Hence, in our Regge limit $C$ is a finite constant.
Finally, the ratio of $u_1$ and $u_2$, which should also be
nonzero and finite in the Regge limit, is given by
$$ u_1/u_2 \ \sim \ Y_2(\theta= - i \pi /4)/Y_2(\theta= i \pi /4)
  + \dots\,. $$
Here we have omitted terms of lower order in the Regge limit.
Using eq.\ \eqref{largemYtwo} we conclude that $m \sin\phi$ has to
be finite, i.e.\ that $\phi$ tends to zero.

For the following analysis of the Regge limit it is convenient to
introduce the parametrization
\be \varepsilon\ =\ \e^{-m \cos \phione} \ \ , \ \ w\ =\ \e^{m
\sin \phione}\ \ , \ \ c\ =\ \cosh(C)\ . \label{eq:xwc} \ee
We shall expand all equations from the previous subsection in the
limit $\varepsilon \rightarrow 0$ while keeping both $w$ and $c$
fixed. From eqs.\ \eqref{ufrompar} we find that
\ba u_1&=&w \varepsilon+O\left(\varepsilon^2\right) \,,
\nonumber \\[2mm]
u_2&=&  w^{-1 }\varepsilon +O\left(\varepsilon^2\right) \,,
\nonumber \\[2mm]
u_3&=& 1- (w+2 c +w^{-1}) \varepsilon
+O\left(\varepsilon^2\right)\ . \ea
One easily verifies that the dependence of the conformal cross
ratios on $\varepsilon$ agrees with their behavior in the Regge
limit, i.e. $u_1,u_2$ and $1-u_3$ vanish in the same order, when
$\varepsilon$ is sent to zero. The leading contributions to all
conformal invariants in the Regge limit are parametrized by the
two functions
$$ \tilde u_1 \ = \ \frac{u_1}{1-u_3} \ \approx \ \frac{w^2}{w^2 + 2cw + 1}
\ \ , \ \ \tilde u_2 \ = \ \frac{u_2}{1-u_3} \ \approx \ \frac{1}{w^2 +
2cw + 1}
\ \ . $$
We can invert these two equations to determine $w$ and $c$ as
functions of $\tilde u_1$ and $\tilde u_2$,
\be w(\tilde u_i) \ \approx \ \sqrt{\frac{\tilde u_1}{ \tilde u_2}} \
\ , \ \ \ c(\tilde u_i) \ \approx \ \frac{1-\tilde u_1 - \tilde
u_2}{2\sqrt{\tilde u_1 \tilde u_2}} \,.
\label{cfromtildeu}
\ee
Our aim now is to evaluate the dependence of the  so-called
remainder function $R$,
\be -R^{(6)}(u_i) \ = \ A^{(6)} - A^{(6)}_{\textrm{div}}-A^{(6)}
_{\textrm{BDS}} \ = \ A^{(6)}_{\textrm{free}}+A^{(6)}
_{\textrm{period}}-\Delta^{(6)}+\mbox{constant}\,, \ee
in the Regge limit as a function of $\tilde u_1$ and $\tilde u_2$.
The function $\Delta$ was spelled out in eq.\ \eqref{eq:Delta}. In
our limit we find that it is given by
\ba - \Delta^{(6)}&=&A^{(6)}_{\textrm{BDS-like}} -
A^{(6)}_{\text{BDS}}\ =\ - \sum_{i=1}^3 \left( \frac1{8}\log^2
u_i+\frac1{4} \text{Li}_2(1-u_i) \right)
\nonumber \\[2mm]
&=& - \frac{1}{4} \log ^2(w) - \frac{1}{4} \log ^2(\varepsilon) -
\frac{\pi ^2}{12} +O\left(\varepsilon\right)\ .
\label{eq:DeltaRegge} \ea
The divergent contribution is actually cancelled  exactly when we
subtract the period contribution to the remainder function
\be A^{(6)}_{\textrm{period}} \ =\ \frac{1}{4}\, m^2 \ = \
\frac{1}{4} \log ^2(w) + \frac{1}{4} \log ^2(\varepsilon)\ .
\label{eq:periodRegge} \ee
It now remains to compute the contribution from the free energy to
leading order in the Regge limit. When we send $m$ to infinity,
the functions $Y_i$ become small all along the real axis so that
we can approximate $\log (1+Y_i(\theta)) \sim Y_i(\theta)$. In this
limit, the free energy then takes the form
\ba A^{(6)}_{\textrm{free}}&\approx& m\int \frac{d \theta}{2 \pi}
\cosh \theta \, Y_{1}(\theta+ i \phione) +m\int \frac{d \theta}{2
\pi} \cosh \theta \, Y_{3}(\theta+ i \phione) \nonumber \\[2mm]  &&
\quad \quad +\sqrt{2} m\int \frac{d \theta}{2 \pi} \cosh \theta \,
Y_{2}(\theta+ i \phione)\ .  \ea
All three integrals can be evaluated easily with the help of the
following integral formula
\be \int_{-\infty}^{\infty} \frac{d \theta}{2 \pi} \cosh \theta
\e^{-a m \cosh \theta} \ =\
 \frac{1}{ \pi} K_1(a m)\ , \ee
such that
\ba A^{(6)}_{\textrm{free}} &\approx& \frac{2 m}{\pi} \cosh (C)
K_1( m) +\frac{\sqrt{2}m}{\pi} K_1(\sqrt{2} m) \ .
\label{AfreeRegge} \ea
In the limit of large $m$ we can use the asymptotic form of
$xK_1(x) \sim \sqrt{\pi x/2} \,  \exp(-x)$. Since  the parameter
$m$ depends on $\varepsilon$ through
$$ m \ = \ \left( \log^2 \varepsilon + \log^2 w \right)^{1/2} \ \approx
- \log \varepsilon - \frac12 \frac{\log^2 w}{\log \epsilon} +
\dots
$$
we conclude that the free energy vanishes when $\varepsilon$ is
sent to zero. Indeed, once we insert the leading dependence of $m
\sim - \log \varepsilon$ in $\varepsilon$ into the formula
\eqref{AfreeRegge} for the free energy, we obtain
\ba
 A^{(6)}_{\textrm{free}} & = & \sqrt{\frac{2}{\pi}}
\ c \ \varepsilon \,  \sqrt{-\log \varepsilon }  + \dots \ .
\label{eq:freeRegge} \ea
where $\dots$ represent lower order terms. Combining this result
with our previous equations \eqref{eq:DeltaRegge},
\eqref{eq:periodRegge} and \eqref{eq:freeRegge} for the Regge
limit of the amplitude $\Delta$ and the periodic contribution we
obtain,
\be - R^{(6)}(u_i) \ \approx \ -\ft{\pi ^2}{12} + \mbox{constant}
+O\left(\varepsilon \right)\ \ . \label{Rphysical} \ee
As we had anticipated, the leading Regge asymptotics of the
remainder function in the physical regime \eqref{physical} is a
constant.

\section{Excited states in thermodynamic Bethe Ansatz}

In this section we shall perform the analytic continuation of the
kinematic variables along the path \eqref{paths}, and we show that
excited states of the system start to contribute. To explain the
basic ideas, we begin the presentation of a toy model
\cite{Dorey:1996re,Dorey:1997rb}. The corresponding analysis of
the system \eqref{NLIE1} and \eqref{NLIE2} is more involved and is
performed in the second subsection.

\subsection{Analytical continuation and excited state TBA}

As a warmup example we want to study a rather simple non-linear
integral equation that depends on a single complex parameter $m$.
We are interested in the behavior of the energy $F_0(m)$ as it is
continued analytically in the parameter $m$. Our toy system
involves a single function $Y(\theta)$ which is determined by
\be -\log Y(\theta) \ =\ m \cosh \theta - \frac{1}{2\pi}
\int_{-\infty}^\infty K(\theta - \theta') \log\left(1 +
Y(\theta')\right)\,.  \label{toyNLIE} \ee
We shall think of the kernel function $K(\theta)$ as being
determined through an S-matrix,
\be  K(\theta)\ =\ -i \frac{\partial}{\partial \theta} \log
S(\theta)  \ . \label{KfromS} \ee
If the S-matrix is unitary, i.e.\ $S(\theta) S^{-1}(-\theta) =1$,
then the function $\log S(\theta)$ is antisymmetric and so
$K(\theta)$ is invariant under $\theta \rightarrow -\theta$.  Let
us furthermore assume that the ground state free energy is given
by
\be F_0(m)\ = \ \frac3{\pi^2} \int_{-\infty}^{\infty} d \theta \,
m \cosh \theta \log\left( 1 + Y(\theta)\right)  . \ee
In order to understand how $F_0(m)$ depends on the mass, it is
quite instructive to study a trivial system with constant
S-matrix. In this case, the equation \eqref{toyNLIE} determines
the solution $\log Y(\theta) = - m \cosh \theta $ and we can
evaluate the free energy of the ground states exactly,
\ba F_0(m) &=& \frac3{\pi^2} \int_{-\infty}^{\infty} d \theta \, m
\cosh \theta \log(1+\e^{-m \cosh \theta}) \ =\ \frac{6 m}{\pi^2}
\sum_{k=1}^{\infty} \frac{(-1)^{k-1}}{k}\, K_1(k m)
\nonumber \\[2mm]
&=& \frac1{2}-\frac{3 m^2}{2 \pi^2} \left[ -\log m+\frac1{2}+\log
\pi - \gamma_E \right] \nonumber \\[2mm]  && +\frac{6}{\pi}
\sum_{k=1}^{\infty} \left( \sqrt{m^2+(2k-1)^2 \pi^2}- (2 k-1) \pi
-\frac{m^2}{2(2k-1) \pi} \right)\ .  \ea
with the Euler-Mascheroni constant ${\gamma}_E=\lim_{s \to
\infty}(\sum_{m=1}^{\infty} m^{-1}-\log s) = 0.577215...$. The
branch cut of the logarithm is chosen to lie along the negative
imaginary axis. For the integration we have used formula (8.526)
from \cite{GrRy}. Then we expressed the infinite sum over Bessel
functions $K_1$ through simpler functions with the help of
\be \sum_{k=1}^{\infty} K_0 (k x)\ =\ \frac1{2}({\gamma}_E+ \log
\ft{x}{4 \pi}) + \pi \sum_{k=1}^{\infty}
\left[\frac{1}{\sqrt{x^2+(2k-1)^2 \pi^2}- \ft1{2 k \pi}} \right]\
. \ee
In addition, we used the formula $[x K_1(x)]'=-x K_0(x)$ for the
derivative of $xK_1(x)$. The free energy $F_0(m)$ possesses a
series of square root singularities at the points $x_k = i(2k-1)
\pi$. Upon analytical continuation along a curve ${\cal C}$ that
encircles the first $N$ of these singularities in the upper half
of the complex plane, we obtain
\be F_0^{{\cal C}}(m)\ =\ F_0(m) - \frac{12}{\pi} \sum_{k=1}
^N\sqrt{m^2+(2k-1)^2 \pi^2} \ . \label{resulttrivial} \ee
We shall reproduce this simple answer down below through an
analysis that does not require the S-matrix to be trivial.

When the system possesses a non-trivial S-matrix, the free energy
can no longer be evaluated exactly. Nevertheless, we can still
analyse how the the free energy $F_0(m)$ changes under analytic
continuation in the mass parameter $m$. We begin by re-writing our
non-linear integral equation \eqref{toyNLIE} in the form
\ba -\log Y(\theta) &=&m \cosh \theta - \frac1{2 \pi}
\int_{-\infty}^{\infty} d \theta' (-i \frac{\partial}{\partial
\theta} \log S(\theta-\theta')) \log(1+ Y(\theta'))
\nonumber \\[2mm]
&=&m \cosh \theta + \frac1{2 \pi} \int_{-\infty}^{\infty} d
\theta' i  \log S(\theta-\theta')
\frac{Y'(\theta')}{1+Y(\theta')}\ .
\ea
Note that the integrand has poles at those points $\theta_0$ at
which the solution $Y(\theta)$ assumes the value $Y(\theta_0) =
-1$. Since the solutions $Y(\theta)$ depend on the parameter $m$,
the solutions of $Y(\theta_0)= -1$ depend on $m$, i.e. they move
through the complex plane as we vary the parameter $m$. Whenever
one of these solutions approaches the integration contour, we
deform it so that solutions never cross. We do that until we get
back to the original value of $m$. At this point we now determine
a solution $Y^{\cal C}(\theta)$ through an equation of the same
form as eq.\ \eqref{toyNLIE} but along some oddly shaped contour
that depends on ${\cal C}$. We want to move the contour back to
the real axis. Along the way we pick up contributions from the
solutions of the equations
\be  Y^{\cal C}(\theta_i)\  = \ - 1 \ \ . \label{toyES1} \ee
Once the integration contour is back on the real axis, our
non-linear integral equation takes the form
\be - \log Y^{{\cal C}}(\theta)\ = \ m \cosh \theta - \frac1{2
\pi} \int_{-\infty}^{\infty} d \theta' K(\theta- \theta')
\log(1+Y^{\cal C}(\theta')) - \sum_{i=1}^{N}(-1)^{n_i} \log
S(\theta-\theta_i)\,, \label{toyES2} \ee
with integers $n_i$ depending on the contour. The two equations
\eqref{toyES1} and \eqref{toyES2} determine the function $Y^{\cal
C}$ along with the parameters $\theta_i$. Similarly, from the
following formula for the free energy
\be F_0(m) \ = \ \frac3{\pi^2} \int_{-\infty}^{\infty} d \theta \,
m \cosh \theta \log(1+ Y(\theta)) \ =\ -\frac3{\pi^2}
\int_{-\infty}^{\infty} d \theta \, m \sinh \theta
\frac{Y'(\theta)}{1+ Y(\theta)} \ . \ee
we infer that
\ba F_0^{\cal C}(m) \ = \ \frac3{\pi^2} \int_{-\infty}^{\infty} d
\theta \, m \cosh  \theta \log(1+ Y^{\cal C}(\theta)) - \frac{6 m
}{\pi} \sum_{i=1}^N (-1)^{n_i} i \sinh  \theta_{i}\,. \label{freeC} \ea
In order to compare with our previous result \eqref{resulttrivial}
for the trivial system, we set $S = 1$ so that $-\log Y^{\cal C} =
-\log Y = m \cosh\theta$. Suppose that $m$ is continued along a
curve in the upper half of the complex $m$-plane that encircles
the points $m= i\pi (2k-1)$ for $k=1, \dots, N$. Then there are
$2N$ solutions of the equation $Y(\theta) = -1$ that cross the
real line. Since these are located at $\pm \theta_k$ with $m \cosh
\theta_k = \pm i(2k-1)\pi$, formula \eqref{freeC} becomes
$$  F^{\cal C}_0(m) \ = \ F_0(m) - \frac{12 m }{\pi}
\sum_{k=1}^N i \sinh  \theta_{k} \ = \ F_0(m) - \frac{12}{\pi}
\sum_{k=1}^N \sqrt{(2k-1)^2\pi^2 + m^2}\ .
$$
Here, we expressed $\sinh\theta_k$ through  $\cosh\theta_k$. The
sum extends over $N$ pairs of solutions $\pm \theta_k$. Note that
both solutions within each pair contribute the same amount.
Therefore, we have simply multiplied the sum over $N$ terms by an
overall factor of two. Our result is identical to our previous
formula for $F_0$ in a trivial system.

We conclude our analysis of the toy model with a few important
comments. The simplest non-trivial case of a continuation occurs
when a single pair $\pm \theta_0$ of solutions to $Y(\theta_0) =
-1$ crosses the real axis. In that case, our equations for
$Y^{\cal C}, F_0^{\cal C}$ and $\theta_0$ become
\ba -\log Y^{{\cal C}}(\theta)\ & = & \ m \cosh \theta + \log
\frac{S(\theta-\theta_0)}{S(\theta + \theta_0)} - \frac1{2 \pi}
\int_{-\infty}^{\infty} d \theta' K(\theta- \theta')
\log(1+Y^{\cal C}(\theta'))\,,   \label{twopole1} \
\\[2mm]
F_0^{\cal C}(m) & = &  \frac{12 m }{\pi}\, i \sinh \theta_{0} +
\frac3{\pi^2} \int_{-\infty}^{\infty} d \theta \, m \cosh  \theta
\log(1+Y^{\cal C}(\theta))\,.
 \label{twopole2} \ea
The position of the $\theta_0$ can be calculated making use of
\ba - \log Y^{\cal C}(\theta_0)& = & i (2k-1) \pi\ = \ m \cosh
\theta_0 - \log S(2 \theta_0) +\log S(0) \nonumber \\[2mm]
& & \hspace*{2cm} - \frac1{2 \pi} \int_{-\infty}^{\infty} d
\theta' K(\theta_0- \theta') \log(1+Y^{\cal C}(\theta'))\,. \ea
The first equality is obtained from eq. \eqref{toyES1} by taking
the logarithm. The second equality just repeats the functional
equation for $Y^{\cal C}$. According to the previous equation, the
right hand side has to stay finite when we send the mass parameter
$m$ to infinity. Since the first term on the right hand side
diverges with $m$, this divergence must be cancelled by one of the
other terms.\footnote{Here we assume that the solution of eq.\
\eqref{toyES1} does not coincide with the zeroes of
$\cosh\theta_0$. This holds true for the cases to be considered in
the next section.}
In the limit $m \rightarrow \infty$, the integral is a small
correction and $S(0)$ does not depend on $m$ at all. Hence, the
term $\log S(\theta_0)$ has to diverge with $m$, i.e.\ the
parameter $2\theta_0$ has to approach one of the poles $\theta_S$
of the bare S-matrix. As we go to finite values of the mass
parameter, this value of $\theta_0$ will receive corrections,
\be \theta_0\ =\ \theta_S/2+ \delta \ . \ee
In principle, we can compute such corrections perturbatively in
the mass parameter $m$. It will turn out later that the case of
two poles crossing the real line is particularly relevant to the
study of the Regge limit.

Before we return to the study of the equations \eqref{NLIE1} and
\eqref{NLIE2}, let us make one comment on the path of the analytic
continuation. In this subsection we assumed the curve ${\cal C}$
to be closed in the complex m-plane, i.e. in the parameter space
of the non-linear integral equation. This is important for the
relation with excited states where $m$ plays the role of  a
physical mass parameter. Our derivations, on the other hand,
remain perfectly valid for arbitrary curves ${\cal C}$ that begin
at $m$ and end at a possibly different point $m'$ of the m-plane.
All we have to do is to replace $m$ by $m'$ in our equation
\eqref{toyES2} and all equations after eq.\ \eqref{freeC}. This
extension shall play an important role in the subsequent analysis.

\subsection{Excited state TBA for six gluon amplitudes}

It is not difficult to extend the analysis of the previous
subsection to the system that describes scattering amplitudes of
six gluons. In this case, we need to solve the equations
\eqref{NLIE1} and \eqref{NLIE2} for three functions $Y_i, i
=1,2,3$. The form of these equations is very similar to eq.\
\eqref{toyNLIE}, with one notable difference.  In the context of
six-gluon amplitudes the physical parameters are the cross ratios
$u_i$ rather than the parameters $m,C,\phione$. Since these two
sets of variables are related by some non-linear equations
\eqref{ufrompar}, an open curve in the space of parameters $m,C$
and $\phi$ may be mapped to a closed curve in the cross ratios.
Consequently, we shall allow our curve ${\cal C}$ to end in a
point $m',C'$ and $\phione'$ that may differ from starting point
$m,C$ and $\phione$. Dealing with the requirement $u_i' = u_i$ is
differed to the next section.

When we analyzed our toy model is was crucial that the kernel
function $K$ could be obtained from an S-matrix through equation
\eqref{KfromS}. In complete analogy, we need to find unitary,
crossing symmetric S-matrices
\be \frac{\partial}{\partial \theta} \log S_j(\theta)\ =\ -2 \pi i
K_j(\theta)\,, \ee
for the two kernels $K_1$ and $K_2$ that were defined in equation
\eqref{Ki}. The solution is given by
\be S_1(\theta)\ = \
   i \frac{1-i e^{\theta }}{1+i e^{\theta }} \,,\quad
 S_2(\theta)\ =\ \frac{2 i \sinh (\theta
   )-\sqrt{2}}{2 i \sinh (\theta )+\sqrt{2}}\ .
\label{Smatrix}
\ee
Up to a common sign, the solution is uniquely fixed by the
unitarity condition
\be S(\theta) S(-\theta)\ =\ S(\theta) S^\star(\theta^*)\ = \  1 \,.
\ee
One may also verify that $S_i$ are crossing symmetric, i.e. that
they obey
\be \bar S_j(\theta)\ =\ S_{a\bar b}(\theta)\ = \ S_{ab}(i \pi
-\theta)\  =\ S_j(i \pi-\theta)\,, \ee
where $\bar S_2(\theta)= S_2(\theta)$ and $\bar S_1(\theta)=
-S_1(\theta)$. Once we have rewritten the functional equations in
\eqref{NLIE1} and \eqref{NLIE2} in terms of $S_i$, we can perform
a partial integration and analyze the behavior of solutions $Y_i$
upon analytic continuation of the parameters $m,C$ and $\phi$. The
results are
\ba - \log Y^{{\cal C}}_2(\theta) & = & -m^\prime \sqrt{2} \cosh
(\theta-i \phione^\prime) - 2 \int_{-\infty} ^{\infty} d \theta'
K_1(\theta-\theta'
) \log (1+ Y^{\cal C}_{2}(\theta')) \nonumber \\[2mm] && \hspace*{-2cm}
- \int_{-\infty} ^{\infty} d \theta' K_2(\theta-\theta') \log ((1+
Y^{\cal C}_{1}(\theta') )(1+ Y^{\cal C}_{3}(\theta'))) +
\sum_{j=1}^{N_1}(-1)^{n_{1,j}} \log S_2(\theta-\theta_{{1,j}})
 \nonumber \\[2mm]  && -2
\sum_{j=1}^{N_2}(-1)^{n_{2,j}}  \log S_1(\theta-\theta_{{2,j}})
-\sum_{j=1}^{N_3}(-1)^{n_{3,j}}  \log
S_2(\theta-\theta_{{3,j}})\,,
\label{eq:ane2}
\ea
and similarly
\ba - \log Y^{\cal C}_{2\pm1}(\theta) & = &
-m^\prime\cosh(\theta-i \phione^\prime) \pm C^\prime -
\int_{-\infty} ^{\infty} d \theta' K_2(\theta-\theta' ) \log (1+
Y^{\cal C}_{2}(\theta')) \nonumber
\\[2mm] && \hspace*{-2cm}
- \int_{-\infty} ^{\infty} d \theta' K_1(\theta-\theta')
\log ((1+ Y^{\cal C}_{1}(\theta'))(1+ Y^{\cal C}_{3}(\theta')) ) \
- \sum_{j=1}^{N_1}(-1)^{n_{1,j}} \log S_1(\theta-\theta_{{1,j}})
\nonumber \\[2mm] && - \sum_{j=1}^{N_2}(-1)^{n_{2,j}}  \log
S_2(\theta-\theta_{{2,j}}) -\sum_{j=1}^{N_3}(-1)^{n_{3,j}} \log
S_1(\theta-\theta_{{3,j}})\,,
\label{eq:ane13}
\ea
with integers $N_s$ and $n_{s,j}$ depending on the path of the
analytical continuation. The parameters $\theta_{s,j}$ are
obtained by solving the equations \be \label{Yiminus1}
 Y^{\cal C}_s (\theta_{s,j}) \ = \ -1 \ \ .
\ee As before, we can now evaluate the formulas \eqref{eq:ane2}
and \eqref{eq:ane13} at the points $\theta_{s,j}$ and insert the
conditions
\be
 - \log Y^{\cal C}_s(\theta_{s,j}) \ = \
i(2k_{s,j}-1)\pi\label{thetasol} \ee
to obtain equations for the positions $\theta_{s,j}$. Since the
resulting expressions are rather lengthy, we shall refrain from
spelling them out. Similarly to the toy model considered in the
previous subsection, the position to the nearest poles in the
limit of large $m$ are related to poles of the S-matrices $S_j$.
Eqs.\ \eqref{thetasol} and \eqref{eq:ane13}, for example, require
that for every $i$ at least one of the differences $\theta_{1,i}-
\theta_{2\pm 1,j}$ or $\theta_{1,i} - \theta_{2,j}$ must approach
a pole of $S_1$ or $S_2$, respectively.

The role of the free energy $F_0$ is played by the functional
$A^{(6)}_{\textrm{free}}$ that was defined in eq.\ \eqref{Afree}.
During the analytical continuation, this quantity changes
according to
\ba A^{(6),{\cal C}}_{free} & = & \int \frac{d \theta}{2 \pi}
m^\prime \cosh \theta \log \left[ (1+Y^{\cal C}_{1}(\theta+ i
\phione^\prime)) (1+Y^{\cal C}_{3}(\theta+ i \phione^\prime))
(1+Y^{\cal C}_{2}(\theta+ i \phione^\prime))^{\sqrt{2}}
\right] \nonumber \\[2mm]
 & & \hspace*{-2cm} -m^\prime i\sum_{j=1}^{\tilde
N_1}(-1)^{\tilde n_{1,j}} \sinh \tilde \theta_{{1,j}} -m^\prime
i\sum_{j=1}^{\tilde N_3}(-1)^{\tilde n_{3,j}}  \sinh \tilde
\theta_{{3,j}} -\sqrt{2}  m^\prime i \sum_{j=1}^{\tilde
N_2}(-1)^{\tilde n_{2,j}} \sinh \tilde \theta_{{2,j}}\,, \label{eq:anA}
\ea
where variables $\tilde \theta_{s,j}$ stand for $
\tilde\theta_{s,j} = \theta_{s,j} - i \phione^\prime$. The
possible $\theta_{s,j}$ that contribute terms to the second line
are solutions of eq.\ \eqref{thetasol} that have passed the line
$\Im \theta = -\phione$ upon analytic continuation. For nonzero
$\phione$, the number of such solutions and the direction in which
they cross the line may differ from the corresponding numbers and
directions for the real line $\Im \theta = 0$. This is why we had
to use new integers $\tilde N_s$ and $\tilde n_{s,j}$ instead of
the corresponding parameters $N_s$ and $n_{s,j}$ that appeared in
the integral equations \eqref{eq:ane2} and \eqref{eq:ane13}. With
these comments we conclude our general discussion of analytic
continuations in the parameters for six gluon scattering. We shall
now return to the Regge limit.

\section{Regge limit of six gluon amplitude}

Now we are prepared to combine the results of the previous
sections in order to complete our main goal, namely to determine
the Regge limit of the six gluon amplitude in the regime
\eqref{newregime}. As explained in section 2, the first step
involves an analytic continuation in u-space. The transformations
\eqref{ufrompar} describe $m,C,\phi$ as a function of the
conformal cross ratios $u_i$ and hence they determine how the
parameters of the integral equations \eqref{NLIE1} and
\eqref{NLIE2} change under analytic continuations along ${\cal
C}_u$. Some important features of the curve in the space of
parameters $m,C,\phi$ are discussed in the first subsection.
Numerical results show that very few solutions of the equations
\eqref{Yiminus1} actually cross the real line while we continue
from the physical to the mixed regime. This enables us to
determine an explicit formula for the Regge limit of the six gluon
amplitude in the regime \eqref{newregime}.

\begin{figure}
\begin{center}
{
}
{\psfrag{vps}{$\varphi/\pi$}
\psfrag{Re}{$\Re m$}
\psfrag{Im}{$\Im m$}
\epsfysize8.cm \epsfbox{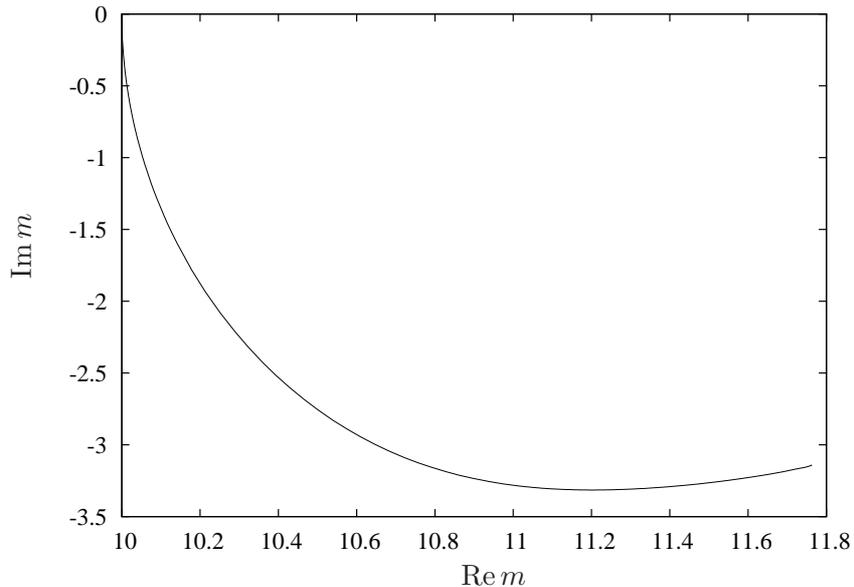}}
\end{center}
\caption{Upon analytic continuation along $\cal C$, the parameter
$m$ is shifted from its initial value. The plot shows numerical
results for $m=10$, $\cosh C= 3/5$ and $\phione=0$ at $\varphi
=0$. Plots for different values of the parameter $C(\varphi=0)$
differ by terms of order $\exp(-m)$.} \label{fig:cpath}
\end{figure}

\subsection{The path of the analytical continuation}
Our first aim is to describe the path of the analytical
continuation. As we explained in section 2, we would like to
continue analytically along the path ${\cal C}_u$ that was defined
in eq.\ \eqref{pathu} with $u_1$ and $u_2$ kept fixed. From the
first two relations in eq. \eqref{ufrompar} we infer that
\be Y_2^- \ \overset{\text{def}}{=} \ Y_{2}(-i\pi/4)\ = \
\frac{u_1}{1-u_1} \,,\quad Y_2^+ \ \overset{\text{def}}{=} \
Y_{2}(i \pi/4) \ =\ \frac{u_2}{1-u_2}\ .  \label{uonetwo} \ee
Note that the left hand side depends on the parameters $m,C,\phi$
which enter the nonlinear integral equations for $Y_i$. Since we
want to keep the right hand side fixed, eqs.\ \eqref{uonetwo}
provide two constraints on the form of the curve in the space of
parameters $m,C,\phi$.

We still need to explore one more relation that involves the cross
ratio $u_3$. Let us recall that $u_3$ has been expressed in eq.\
\eqref{uYrelations2} through the values of the functions $Y_i$ in
the strip $|{\mbox{\it Im\/}}(\theta-i \phione)|< \pi/2$. We can
actually turn eq.\ \eqref{uYrelations2} into a formula for $C =
C(u_i)$ once we notice that
\be Y_{3}(\theta)/Y_{1}(\theta)\ = \ Y_{3}(0)/Y_{1}(0)\ = \
\e^{2C} \ . \label{YthreeYone} \ee
This simple relation follows by subtracting the non-linear
integral equations \eqref{NLIE2} for $Y_1$ and $Y_3$ from each
other. We employ eq.\ \eqref{YthreeYone} to eliminate both $Y_{3}$
and $Y_{1}$ from the equation for $u_3$ and arrive at,
\be \cosh^2 C \ =\ \frac{(1- u_1-u_2-u_3)^2}{4 u_1 u_2 u_3} \ .
\label{uthree} \ee

During our continuation we keep $u_1$ and $u_2$ fixed while $u_3$
is moved along a full circle until it comes back to its original
position. Because the left hand side involves the square of $c =
\cosh C$, the quantity $\cosh C$ has changed its sign once we
reach the end-point of the curve ${\cal C}_u$. Note that eq.\
\eqref{uthree} determines the parameter $C$ explicitly as a
function of the cross ratios. The variation of the real and
imaginary part of $C$ along the path ${\cal C}_u$ is plotted in
fig. 3. All the curves start in the lower half on the left hand
side and proceed to the central line at $Im C = \pi/2$ and return
to the left and into the upper half of the figure. At $\varphi =
\pi/2$, the real part of $C$ is of order $m$.

The dependence of $m$ and $\phione$ on $\varphi$ is a little more
difficult to obtain. It requires inverting the eqs.\
\eqref{uonetwo} which contain the function $Y_2$. The latter must
be determined numerically by solving the non-linear integral
equation for a large number of values $m,\phione$ and $C$. Given
$u_i$ we can restrict the values of $C$ to
$$\cosh^2 C(\varphi) \ =\ \frac{(1- u_1-u_2-u_3 e^{-2i\varphi})^2}
{4 u_1 u_2 u_3 e^{-2i\varphi} }\,,$$
where $\varphi$ runs from $\varphi =0$ to $\varphi = \pi$. We have
performed these numerical studies for special cases in which $u_1
= u_2$. The condition amounts to setting $\phione =0$ at $\varphi
=0$. It is not difficult to see that $\phione =0$ will remain true
as we move along the contour ${\cal C}_u$. Hence, it remains to
determine $m(\varphi)$ such that $Y_2(\pm i \pi /4)$, and hence
the cross ratios $u_1 = u_2$, remain invariant as me change
$\varphi$. An example is shown in fig. 2. The shape of the curve
has very little dependence on the parameter $C$. We observe that
the parameter $m$ becomes complex. At $\varphi = \pi$, its
imaginary part is $\Im m(\pi) = -\pi$. The real part also gets
shifted by an amount $\Delta m = \Re m(\pi)-\Re m(0) \sim 1.8$. We
shall reproduce these shifts through an analytical argument in the
next section. Some more comments on the numerical studies can be
found in appendix C.

\begin{figure}
\begin{center}
{
}
{
\psfrag{ReC}{$\Re C$}
\psfrag{ImC}{$\Im C$}
\psfrag{c=0.0}{\!\!\!\!\!\! $c = 0.0$}
\psfrag{c=0.3}{\!\!\!\!\!\! $c = 0.3$}
\psfrag{c=0.6}{\!\!\!\!\!\! $c = 0.6$}
\psfrag{c=1.0}{\!\!\!\!\!\! $c = 1.0$}
\psfrag{c=2.0}{\!\!\!\!\!\! $c = 2.0$}
\epsfysize8.cm \epsfbox{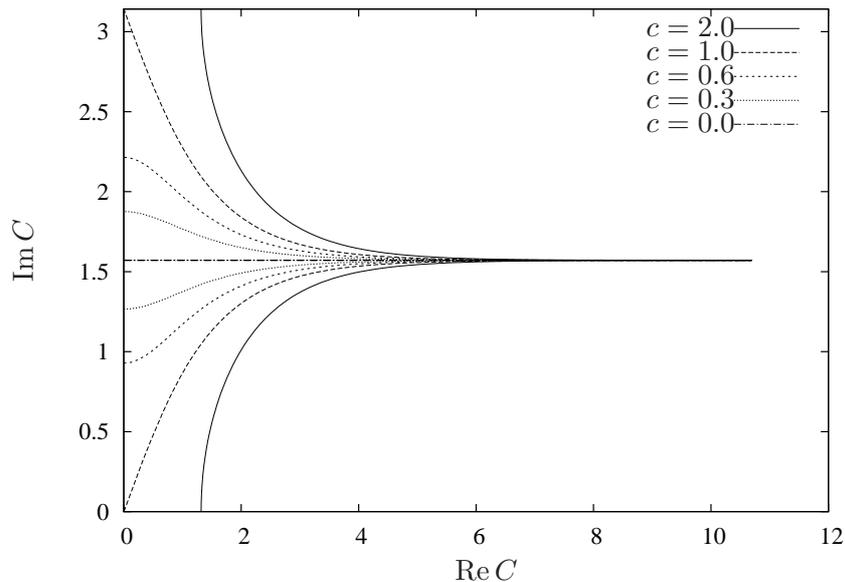}}
\end{center}
\caption{The plot shows the dependence of $C$ on $\varphi$. The
various curves correspond to the values $m=10$, $\phi=0$ and
$\cosh C=c$ at the starting point $\varphi=0$. Here, we plot $C$
in its complex plane along the path given by \eqref{uthree} while
$\phi(\varphi)=0$.} \label{fig:cpath2}
\end{figure}

\subsection{Evaluation of the Regge limit}

In the previous section we have investigated the image of curve
${\cal C}_u$ in the space of parameters $m,\phione$ and $C$. For
$\phione = 0$, the corresponding values of $m(\varphi)$ and
$C(\varphi)$ are depicted in figs.\ 2 and 3. These numerical data
provide us with a good qualitative understanding of the curve
along which we would like to continue the parameters of the
non-linear integral equations \eqref{NLIE1} and \eqref{NLIE2}. As
we discussed in section 4, we need to watch the solutions of
$Y_i(\theta) = -1$ while continuing $m,\phione$ and $C$. Our
numerical calculations show that, for large values of $m$, very
few poles get close to the integration contour. In the case of
$Y_3$, the numerical results are displayed in fig.\ 4. The plot
was produced for a continuation along the paths described by figs.
2 and 3. We see a single pair of poles that crosses the real line.
The symmetry under reflection $\theta \rightarrow - \theta$ is a
direct consequence of our special choice $\phione =0$. Analogous
results for the solutions of $Y_2 = -1$ are displayed in fig. 5.
In this case, one pair of solutions approaches the real line while
all others stay away from it. We did not display the corresponding
plot for $Y_1$ since it looks very similar to that for $Y_3$,
except for an overall shift by $\pm i \pi$.

With these results in mind we can actually spell out a system of
equations that should determine the solutions $Y_i$ at the end of
the analytic continuation, i.e.\ in the mixed regime,
 \ba - \log Y^\prime_2(\theta) & = & -m^\prime \sqrt{2}
\cosh (\theta-i \phione^\prime)  - \int_{-\infty} ^{\infty} d
\theta'
K_2(\theta-\theta') \log ((1+ Y^\prime _{1}(\theta') )
(1+ Y^\prime _{3}(\theta'))) \nonumber \\[2mm] && \hspace*{-1cm}
- 2 \int_{-\infty} ^{\infty} d \theta' K_1(\theta-\theta' ) \log
(1+ Y^\prime _{2}(\theta')) + \log S_2(\theta-\theta_1) -  \log
S_2(\theta-\theta_2)\,,\label{eq:ane2twopoles} \\[2mm]
- \log Y^\prime _{2\pm1}(\theta) & = & -m^\prime\cosh(\theta-i
\phione^\prime) \pm C^\prime - \int_{-\infty} ^{\infty} d \theta'
K_1(\theta-\theta') \log ((1+ Y^\prime _{1}(\theta'))(1+ Y^\prime
_{3}(\theta')) ) \nonumber
\\[2mm] && \hspace*{-1cm}-
\int_{-\infty} ^{\infty} d \theta' K_2(\theta-\theta' ) \log (1+
Y^\prime _{2}(\theta'))  \ + \log S_1(\theta-\theta_1) - \log
S_1(\theta-\theta_2)\ .\label{eq:ane13twopoles} \ea
One might have expected to see additional terms from the solutions
of $Y_2 = -1$ on the right hand side. As one may easily see, these
take the form
\be \lim_{a \rightarrow 0} \left(2 \log(S_1(\theta - a)) - 2
\log(S_1(\theta + a))\right) \ = \  \lim_{a \rightarrow 0} \left(
4 i \sech(\theta) a\right) \ = \ 0\,, \ee and
\be \lim_{a \rightarrow 0} \left(2 \log(S_2(\theta - a)) - 2
\log(S_2(\theta + a))\right) \ = \  \lim_{a \rightarrow 0} \left(
8 i \sqrt{2} \cosh(\theta) \sech(2\theta) a\right) \ = \ 0 \ .\ee
Hence, we there are no additional terms on the right hand side of
eqs.\ \eqref{eq:ane2twopoles} and \eqref{eq:ane13twopoles}. In the
above equations and throughout the entire discussion below we
place a $\prime$ on all quantities that have been continued to the
endpoint of the curve ${\cal C} = {\cal C}_u$ that we described in
the previous subsection. The parameters $\theta_1$ and $\theta_2$
appearing in the arguments of the two S-matrices \eqref{Smatrix}
must solve one of the equations
\be \label{thetaonetwo}
 Y'_1(\theta_\nu) \ = \ -1 \,, \quad
 Y'_3(\theta_\nu) \ = \ -1
\
\ee
depending on value of $C'$.
In addition we shall use the convention that, after crossing the
contour, $\theta_{1}$ has positive imaginary part while
$\theta_{2}$ has negative imaginary part. For $\phione=0$ we have
$\theta_{1}=-\theta_{2}$.

\begin{figure}
\begin{center}
\begin{picture}(360,240)
\put(180,-10){$\Re \theta$}
\put(-10,120){\rotatebox{90}{$\Im \theta$}}
\put(5,5){\epsfysize8.0cm \epsfbox{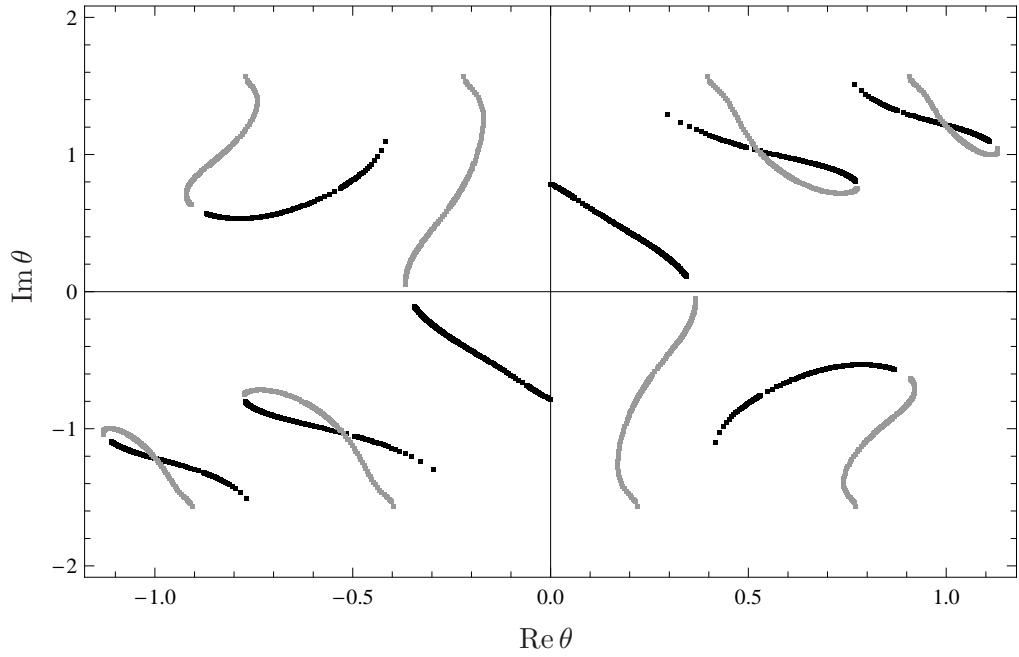}}
\end{picture}
\end{center}
\caption{Positions of some central solutions to the equation
$Y_3(\theta)=-1$ along the path depicted in figs.~\ref{fig:cpath},
\ref{fig:cpath2}. Light dots correspond to positions in the first
half of the path before any solution has crossed the real axis
while the dark ones are associated with the second part.}
\label{fig:poles1}
\end{figure}

\begin{figure}
\begin{center}
\begin{picture}(360,240)
\put(180,-10){$\Re \theta$} \put(-10,120){\rotatebox{90}{$\Im
\theta$}} \put(5,5){\epsfysize8.0cm \epsfbox{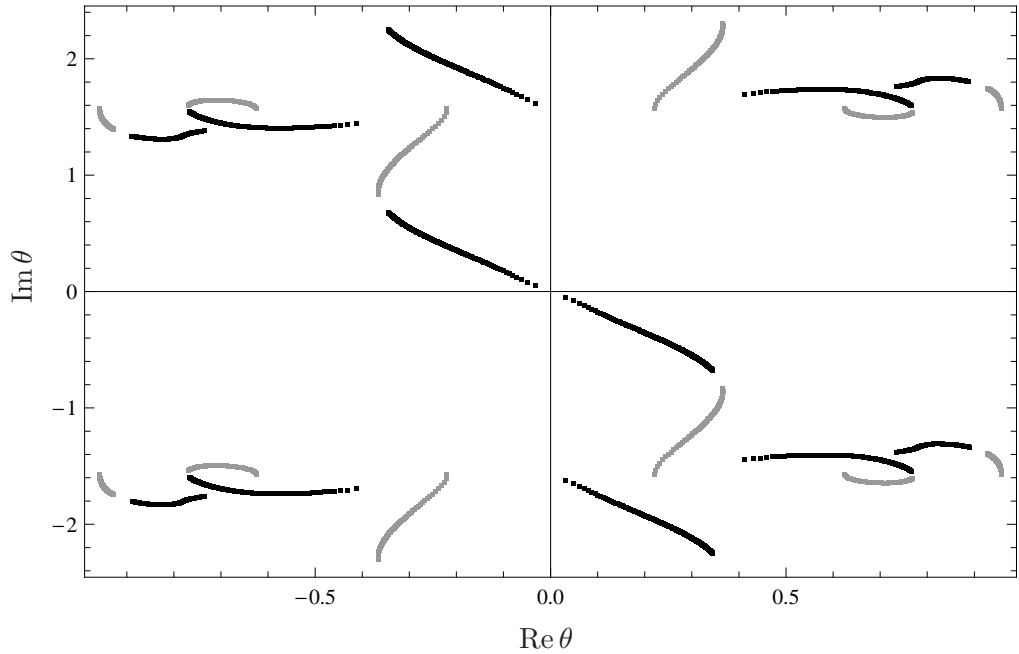}}
\end{picture}
\end{center}
\caption{Positions of some central solutions to the equation
$Y_2(\theta)=-1$ along the path depicted in figs.~\ref{fig:cpath},
\ref{fig:cpath2}. We change from light to dark dots at the point
where the first solutions of $Y_{3}= -1$ cross the real axis.}
\label{fig:poles2}
\end{figure}

In the leading approximation for large values of the mass
parameter $m$, the non-linear integral equations simplify,
\ba \log Y^\prime_{2}(\theta) & =& -m^\prime \sqrt{2} \cosh
(\theta-i \phione^\prime) -\log S_2(\theta-\theta_{1}) +\log
S_2(\theta-\theta_{2})\,,
\nonumber \\[2mm]
\log Y^\prime_{2\pm1}(\theta)&=& -m^\prime\cosh(\theta-i
\phione^\prime) \pm C^\prime -\log S_1(\theta-\theta_{1}) +\log
S_1(\theta-\theta_{2}) \ .  \ea
As we explained earlier, the solutions to eq.\ \eqref{thetaonetwo}
are determined by the poles of the S-matrix, at least to leading
order in large $m$. Since $S_2$ has poles at $\theta_S$ with $\exp
i \theta_S = i$, the solutions $\theta_\nu \sim \theta_S/2$ are
well approximated by
 \be \theta_{1}\ =\ -\theta_{2}\ =\ i \ft{\pi}{4}+\ldots\,,  \ee
with corrections of order $\exp(-m)$. We can insert the formulas
for $Y_i^\prime$ with $\theta_1 = - \theta_2 = i\pi/4$ into the
equations \eqref{uonetwo} and \eqref{uthree} that determine the
conformal cross ratios $u'_i$ at the end-point of the analytic
continuation. A short computation gives
\ba u_1^\prime&=&  w^\prime \gamma \varepsilon
^\prime+O\left((\varepsilon ^\prime)^2\right)\,,
\nonumber  \\[2mm]
u_2^\prime&=& (w^\prime)^{-1} \gamma \varepsilon^\prime
+O\left((\varepsilon ^\prime)^2\right)\,,
\label{uprime}  \\[2mm]
u_3^\prime&=&1 -  \left( w^\prime-2 c^\prime +
(w^\prime)^{-1}\right) \gamma
   \varepsilon^\prime +O\left((\varepsilon ^\prime)^2\right)\,,
\nonumber\ea
with $\gamma = - (3 + 2 \sqrt 2)$. Here, the parameters
$w',c^\prime$ and $\varepsilon^\prime$ are determined by our
formulas \eqref{eq:xwc} only that we have to replace $m,\phi,C$ by
$m',\phi',C'$. At the end of the analytic continuation we want to
return to the original values of the cross rations, i.e.\ we need
to impose $u_i = u_i^\prime$. This leads to
\be \varepsilon ^\prime\ =\  \gamma^{-1} \varepsilon
+O\left(\varepsilon ^2\right)\,,\quad w^\prime\ =\
w+O\left(\varepsilon \right)\,,\quad c^\prime\ =\
-c+O\left(\varepsilon \right)\,,\quad \label{wecprime} \ee
Obviously, there exists a second solution of the conditions $u_i =
u_i^\prime$. This solution, however,  violates the condition $c =
- c^\prime$ that we have argued for in the previous subsection.

It is instructive to re-express our relations \eqref{wecprime}
through the parameters $m,C$ and $\phi$. We have seen already that
$\cosh C' = - \cosh C$. For the other two parameters we obtain
\ba & & m^\prime \ =\  \sqrt{m^2+s_h^2 +2 m s_h \cos\phione}
+O\left(\e^{-m}\right) \,, \label{mprimem}
\\[2mm]
& & m^\prime \cos \phione^\prime\ = \ s_h+m \cos \phione
+O\left(\e^{-m}\right)\,, \label{phiprimephi} \ea
where $ s_h= \log \gamma= -i\pi + \log(3+2\sqrt2)$. Let us recall
that these relations between the value of $m, C$ and $\phi$ in the
initial and final point of our curve ${\cal C}$ were derived under
the assumption of single pole crossing. The numeric results we
displayed in the previous section were obtained for $\phione = 0$.
In this case, the assumption of single pole crossing gives
$\phione' = 0$ and
\ba m^\prime \ = \ m- i \pi +\log(3+ 2
\sqrt{2}))+O\left(\e^{-m}\right)\, .  \ea
This coincides with the shift we found in our numerical studies,
see fig. 2, and provides indirect evidence for the assumption made
in the analytic treatment.

Let us now turn to the main goal, namely to the calculation of the
remainder function
\be -R^{\prime(6)}\ = \ A^{\prime(6)}_{\textrm{free}
}+A^{\prime(6)}_{\textrm{period}
}-\Delta^{\prime(6)}+\mbox{constant}\,, \ee
where the individual terms are not evaluated after analytic
continuation along the curve ${\cal C}$. Let us begin with the
function $\Delta^\prime$ that was spelled out in eq.\
\eqref{eq:Delta}. Upon continuation along the curve \eqref{pathu}
we find
\ba \Delta^{\prime(6)}&=&\Delta^{(6)}-\ft{\pi i}{2} (\log u_3 -
\log(1-u_3))-\ft{\pi^2}{2} \nonumber \\[2mm]
&=& \ft{i}{2} \pi  \log \varepsilon + \ft{1}{2} i \pi \log \left(2
c+w+ w^{-1}\right) -\frac{\pi^2}{2} +\Delta^{(6)}
+O\left(\varepsilon \right)\ . \ea
In passing from the first to the second line we have inserted the
asymptotic form \eqref{uprime} of the cross ratios $u_i^\prime =
u_i$. The period contribution is even easier to find,
\ba A^{\prime(6)}_{\textrm{period} }&=& \ft1{4} (m^\prime)^2\ = \
\ft1{4} (m^\prime)^2 -\ft1{4} (m)^2 +A^{(6)}_{\textrm{period} }
\nonumber \\[2mm]
&=& - \frac12 \log\gamma \, \log(\varepsilon )+ \frac{1}{4} \log^2
\gamma + A^{(6)}_{\textrm{period} } +O\left(\varepsilon \right)\nonumber\\
&=&  \ft{i}{2} \pi \log(\varepsilon) - \frac12 \log|\gamma| \, \log(\varepsilon )+ \frac{1}{4} \log^2
\gamma + A^{(6)}_{\textrm{period} } +O\left(\varepsilon \right)\,.
\ea
Finally, we also have to determine the contribution from the free
energy. Specializing our general result \eqref{eq:anA} to the
present case we find that
\ba A^{\prime(6)}_{\textrm{free}} & = & \int \frac{d \theta}{2
\pi} m^\prime \cosh \theta \log \left[ (1+Y^\prime_{1}(\theta+ i
\phione^\prime)) (1+Y^\prime_{3}(\theta+ i \phione^\prime))
(1+Y^\prime_{2}(\theta+ i \phione^\prime))^{\sqrt{2}}
\right] \nonumber \\[2mm]
 & & \hspace*{2cm} -m^\prime i \sinh (\theta_1 - i\phi')
 +m^\prime
i \sinh (\theta_2 - i\phi')+ \sqrt{2} m' \sinh(|\phi'|)\,.
\label{Afree1} \ea
Note that the second line contains a contribution from one
solution of the equation $Y_2 = -1$. As we explained at the end of
section 4.2, the contributions to the free energy arise from
solutions $\theta_*$ that cross the line $\Im \theta_* = \phione$.
As one can read off from fig. 5, one of the solutions of $Y_2 = -1$
does cross this line, at least when $\phione \neq 0$.

In the limit of large $m$, the integral in eq.\ \eqref{Afree1}
gives a vanishing contribution as one may see by repeating the
analysis we have carried out in great detail in section 3.2. So
the only terms that are relevant in the Regge limit come from the
second line of the previous equation. This gives
\ba A^{\prime(6)}_{\textrm{free} } &=& \sqrt{2} m^\prime
\cos(\phione^\prime) + \sqrt{2} m' \sinh(|\phi'|) +
O\left(\varepsilon \right) \ = \ \sqrt{2} \log \varepsilon' +
\sqrt{2} |\log w'|
 + O\left(\varepsilon \right)\nonumber \\[2mm]
&=& - \sqrt{2}\log\varepsilon + \sqrt{2}\log \gamma +
A^{(6)}_{\textrm{free}} + \sqrt{2} |\log w|+O\left(\varepsilon
\right)\,. \ea
In passing to lower line, we have inserted eq. \eqref{phiprimephi}
and also added the term $A_{\textrm{free}}$ that was evaluated in
section 3.2. Since the free energy before analytic continuation
vanishes in the limit of $\varepsilon \rightarrow 0$, the
additional term is actually zero. We only put it back in to show
that the leading term should be regarded as a difference of
energies. Using the final result \eqref{Rphysical} of our
computation in section 3.2, we are now prepared to determine the
Regge limit of the remainder function in the regime
\eqref{newregime},
\ba -R^{\prime(6)}&=& A^{\prime(6)}_{\textrm{free}
}+A^{\prime(6)}_{\textrm{period} }-\Delta^{\prime(6)}
\nonumber \\[2mm]
&=& -  e_2 \log \varepsilon  -\ft{\pi}{2} i \log \left(2
c+w+w^{-1}\right) + \sqrt{2} |\log w| + \mbox{constant}^\prime
+O\left(\varepsilon \right)\,. \ea
Here, we have moved all the constant terms into a redefinition of
the `constant' and defined the constant $e_2$ by
\be
e_2 \ = \ \left(\sqrt{2}+ \ft{1}{2} \log (3+2 \sqrt{2}) \right)\,.
\ee
In a final step, we want to express the variables $w,c$ and
$\varepsilon$ through
the conformal cross ratios. Using eqs.\
\eqref{cfromtildeu} along
with
\be
\varepsilon \ =\
\sqrt{\tilde u_1 \tilde u_2} (1- u_3)\,,
\ee
we find that
\ba -R^\prime &=& - e_2 \log((1-u_3)\sqrt{\tilde u_1 \tilde
u_2}) +\ft{1}{2} i \pi \log \sqrt{\tilde u_1 \tilde u_2} +
\sqrt{2} |\log( \tilde u_1/ \tilde u_2)| \nonumber \\ &&+\mbox{const}^\prime +
O\left(1- \tilde u_3\right) \ . \ea
After exponentiation of this result we obtain
\ba \mbox{Amp}^\prime&\sim& \langle W^\prime \rangle\sim
\exp\left[-\frac{\sqrt{\lambda}}{2 \pi} A^\prime\right]
=\exp\left[-\frac{\sqrt{\lambda}}{2 \pi}
(A_{\textrm{div}}^\prime+A_{\textrm{BDS}}^\prime-R^\prime)\right]\,,
\ea
with
%
\begin{multline}
\e^{\frac{\sqrt{\lambda}}{2 \pi} R^\prime} \ \sim
\ e^{-i \ft{\pi}{2} \frac{\sqrt{\lambda}}{4 \pi} \ln (\tilde u_1 \tilde u_2)}
\left(
(1-u_3) \sqrt{\tilde u_1 \tilde u_2}  \right)
^{\frac{\sqrt{\lambda}}{2 \pi} e_2}
\\
\times \left[
\left(\frac{\tilde u_1}{\tilde u_2} \right)^{-\frac{\sqrt{\lambda}}{\sqrt{2} \pi}}
\theta\left(\frac{\tilde u_1}{\tilde u_2} - 1\right)
+ \left(\frac{\tilde u_2}{\tilde u_1} \right)^{-\frac{\sqrt{\lambda}}{\sqrt{2} \pi}}
\theta\left(\frac{\tilde u_2}{\tilde u_1} - 1\right) \right]
\,,
\label{leadingstrong}
\end{multline}
up to some irrelevant constant
prefactor. This formula is the main result of our analysis. It
describes the Regge limit of the six-gluon amplitude in the mixed
regime and at strong coupling. As we have stressed before, our
calculation assumes that the analytical continuation of the
amplitude may be performed within the prescription of Alday et
al., see our discussion in the introduction and the next section.

Before we conclude, we would like to compare our result with the
form of the answer at weak coupling. The relevant expression is
known in the leading logarithmic approximation
\cite{Bartels:2008sc}. The following formula gives the correction
to the BDS expression:
\ba
\mbox{Amp}' = \mbox{Amp}_{2 \to 4}^{BDS} \,\,\left( 1+ i \Delta_{2 \to 4} \right)\,,
\label{BLV1}
\ea
with
\ba \Delta_{2 \to 4} \ =\ \frac{a}{2}\sum_{n=-\infty}^{n=\infty}
(-1)^n \int \frac{d \nu}{\nu^2 + \frac{n^2}{4}}
(s_2^{\omega(\nu,n)}-1) \left(\frac{q_2^*p_4^*}{p_5^*
q_1^*}\right)^{i\nu -\frac{n}{2}} \left(\frac{q_2 p_4}{p_5
q_1}\right)^{i\nu +\frac{n}{2}} \ .
\label{BLV2}
\ea
All relevant notations are explained in \cite{Bartels:2008sc}. In
particular, the $q_i$ are the transverse momenta of the $t_i$
invariants, the parameter $a$ is given by $a=\lambda/8\pi^2$ and
$\lambda$ is the 't Hooft coupling $\lambda = g^2 N_c$. The
exponent $\omega$ of $s_2$ is given by
\ba \omega(\nu,n)\ =\ 4a{\cal R} \left( 2 \psi(1) - \psi(1+i \nu
+\frac{n}{2}) - \psi(1+i \nu -\frac{n}{2}) \right)\ . \ea
In the weak coupling expression \eqref{BLV1} the new term, $\Delta_{2 \to 4}$,
comes as an additive correction. There is no indication
that this correction should be written as an exponential.
From the point of view of Regge theory, it seems more natural to sum over
the different singularities in the complex angular momentum plane, and
it is not clear whether exponentiation would still be consistent
with Regge factorization. However, the leading logarithmic approximation alone
is not sufficient to exclude the possibility of exponentiation.

The function $\Delta_{2 \to 4}$ may be re-expressed in terms of the
cross ratios $u_i$ that we have built in section 2. In these
variables, \eqref{BLV2} assumes the form
\ba \Delta_{2 \to 4}\  = \ \frac{a}{2} \sum_{n=0}^{n=\infty}
(-1)^n \int \frac{d \nu}{\nu^2 + \frac{n^2}{4}} ((1 -
u_3)^{-\omega(\nu,n)}-1) \left(\frac{\tilde u_1}{\tilde u_2}\right)^{i\nu} \cosh n C\,,
\label{BLV3}
\ea
where we have used $1-u_3 \sim 1/s_2$ (appendix A),
\be
\Big|\frac{q_2p_4}{p_5q_1}\Big|= \sqrt{\frac{\tilde u_1}{\tilde u_2}} \approx w \,\,,
\ee
and
\ba
\cos \left(\arg\frac{q_2p_4}{p_5q_1}\right) = \frac{1 - u_1 - u_2 - u_3}{2\sqrt{u_1u_2}}
\approx \frac{1 - u_1 - u_2 - u_3}{2\sqrt{u_1u_2u_3}} = \cosh C . \ea

The leading power of $s_2$ belongs to $n=1$ and $\nu =0$ so that
the intercept at weak coupling becomes
\ba \omega(0,1)\ =\  - E_2 \ = \ \frac{\lambda}{\pi^2}(2 \ln 2 -
1)\ . \ea
Consequently, up to logarithmic corrections, the leading
asymptotic behavior of the remainder function at weak coupling is
given by
 \ba \Delta_{2 \to 4} \ \sim\  \frac{a}{2} (1 -
u_3)^{-\omega(0,1)}\,\, \cosh C\,. \label{leadingweak} \ea
In this leading logarithmic approximation, it is not possible to
determine the scale of the energy $s_2$. Put differently, the
factor $1-u_3$ may be accompanied by some undetermined finite
function of $\tilde u_1$ and $\tilde u_2$.

Let us attempt a first comparison of our strong coupling result
eq.\ \eqref{leadingstrong}
with the weak coupling amplitude in eqs.\ \eqref{BLV1},\eqref{BLV2} and
eq.\ \eqref{leadingweak}. In both cases, the energy dependence of
the correction is power-behaved (i.e. Regge-behaved).
We first face the fact that, at weak coupling, the correction is additive
whereas, at strong coupling, the remainder function in the
exponent leads to a multiplicative correction to the BDS
amplitude. As a possible interpretation, we might assume
that in eq.\ \eqref{BLV1} the correction, which at weak coupling is of
order $\lambda$, grows with $\lambda$ and, at strong coupling,
dominates over the $1$. We are then led to confront $\Delta_{2 \to 4}$
in eqs.\ \eqref{BLV3} and \eqref{leadingweak} with eq.\ \eqref{leadingstrong}.

A striking observation is the change in the sign of the exponent
$E_2$: The intercept of the energy $s_2$ which on the weak
coupling side is positive appears to turn negative when the
coupling is increased. Also, the strong coupling expression
exhibits an energy scale depending upon the product of $\tilde
u_1$ and $\tilde u_2$. Next let us look at the ratio $\tilde
u_1/\tilde u_2$ which is raised to power $i \nu$ on the weak
coupling side. Since at large $s_2$ the leading behavior comes
from $i \nu=0$, the dependence on $\tilde u_1/\tilde u_2$ drops
out of eq.\ \eqref{leadingweak}. In contrast to this, on the
strong coupling side, the ratio $\tilde u_1/\tilde u_2$ comes with
an exponent of order $\sqrt{\lambda}$. In addition, the factor
$\cosh C$ that multiplies the answer at weak coupling, is absent
at strong coupling. Finally, in eq.\ \eqref{leadingstrong}, we see
a phase factor which also depends upon the product of $\tilde u_1$
and $\tilde u_2$. It should be kept in mind, however, that the
full BDS amplitude at strong coupling, $\ A^{(6)}_{\textrm{div}} +
A^{(6)}_{\textrm{BDS}}$, when continued into the mixed region,
also contains a phase: A straightforward calculation shows that
this phase exactly cancels the phase in eq.\
\eqref{leadingstrong}. As a result, the strong coupling result has
no such phase factor.\footnote{In the weak coupling regime, it was
the appearance of such a phase factor in the BDS amplitude which
played a vital role in understanding the deficiency of the BDS
formula. The complete cancellation of the phases at strong
coupling therefore provides support for the correct analytic
structure at strong coupling.}

Let us try a further interpretation and assume that, at strong
coupling, the correction factor $\exp(\sqrt{\lambda}R'/{2\pi})$
can also be written in the form of eq. \eqref{BLV3} as an
expansion in the two-dimensional conformal group (M\"obius group),
with some strong coupling analogue $\tilde \omega(\nu,n)$ of the
function $\omega(\nu,n)$. Then, in eq.\ \eqref{leadingstrong}, the
exponent of $\tilde u_1/\tilde u_2$ could be interpreted as coming
from a dominant $\nu$-value $\nu_{s}$ that is of the order
$\sqrt{\lambda}$. At the same time, the exponent of $s_2$ would
come from the function $\tilde \omega(\nu,n)$, evaluated at the
value $\nu_s$. Finally, the absence of the factor $\cosh C$ might
result from the fact that, at large $\nu_s$, $\tilde
\omega(\nu_s,0)$ dominates over $\tilde \omega(\nu_s,1)$. In
pursuing such an interpretation one would perform a saddle point
analysis of the $\nu$-integral: At weak coupling, the large
parameter is $\ln s_2$ which leads to the dominance of the point
$\nu =0$. In contrast, at strong coupling it is $\sqrt{\lambda}$
which plays the role of the large parameter, leading to a value
$\nu_2 = \cal O(\sqrt{\lambda})$. In this sense, the form of our
strong coupling results seems perfectly consistent with the
general features of Regge theory.

\section{Conclusions and Outlook}

In this paper we have studied the Regge limit of the planar six-point
function at strong coupling. In the weak coupling regime this high
energy limit, taken in a very special kinematic region ('mixed region')
where some energies are positive while others are negative, is known to
be particularly sensitive to the analytic properties of scattering
amplitudes. Making use of the set of equations for $Y$-functions,
we have been able to perform the analytic continuation of the
scattering amplitude at strong coupling into this kinematic region
and to calculate new corrections which are given by excitations of
the auxiliary quantum integrable system. A comparison of the strong
coupling result with the expressions for weak coupling shows
interesting features but needs further investigations. At the
moment, nothing is known about how the quantity $E_2$ may be
interpolated between weak and strong coupling.

As we pointed out in the introduction, our calculation of the
parameter $E_2$ at strong coupling was based on one important
assumption, namely that we can simply continue the prescription of
Alday et al. into the mixed regime. In other words, we have
assumed that the amplitude in the mixed regime is determined by
the exponential of the free energy in the auxiliary quantum
integrable system. One may certainly wonder whether this
assumption was justified. As we sketched in the introduction,
scattering amplitudes at strong coupling do possess a geometric
origin. According to the original prescription of Alday and
Maldacena, they are determined by the area of a 2-dimensional
surface $S_6$ which stretches to the boundary of $AdS_5$. This
surface emerges as the unique saddle point that dominates the
string theoretic path integral in the limit of strong 't Hooft
coupling. The kinematic invariants of the scattering process are
encoded in the shape of the 1-dimensional boundary $\partial S_6$
of $S_6$ in the boundary of $AdS_5$. Our analytic continuation in
the space of kinematic invariants amounts to a continuation in the
boundary conditions for a (path) integral. In any saddle point
approximation, the contributing saddle points can be very
sensitive to boundary conditions. In fact, there could even exist
points, i.e. values of the kinematic invariants in our case, at
which several saddle points contribute to the same order. From
this point of view it seems far from obvious that amplitudes in
the mixed regime can be obtained by simple analytic continuation
and that they contain a single exponential term. This issue
certainly deserves further investigation.

There are a few more or less obvious extensions of this work. It
seems worthwhile exploring the relation between the analytic
structure of scattering amplitudes and excited states of the
auxiliary quantum system for more general cases, including other
regimes and for a larger number of external gluons. In this
context it might be useful to work with an alternative version of
the non-linear integrable equations \cite{Alday:2010ku} in which
only the geometric cross ratios appear as fundamental
parameters.\footnote{Since the contribution $A^{(6)}_\textrm{period}
$ is most easily expressed in terms of the variable $m$, however,
one must still determine this parameter as a function of the cross
ratios $u_i$.} Another question addresses the relation between
excited states and analytic continuation. For other quantum
integrable systems it is well-known that there exist excited
states that are not related to analytic continuation in the
parameters. The associated integral equations are of the same
form, but without any constraints in the numbers $N$ of driving
terms. One may wonder whether such excited states could also have
some interpretation in the context of scattering amplitudes.

\bigskip

\noindent
{\bf{Acknowledgements:}} We thank Lev Lipatov, J\"org Teschner and
Pedro Vieira for interesting discussions and comments on the
manuscript. VS is grateful to the String Theory group at the UBC
Vancouver and to the Pacific Institute of Theoretical Physics
for their hospitality during the final stages of the project.
This work was supported by the SFB 676.

\appendix
\section{Kinematics in the Regge limit}
\label{ap:rl}

In this appendix we shall analyze the behavior of the conformal
cross ratios $u_i$ in the ordinary multi-Regge limit
\be s \ \gg \ s_1, s_2, s_3 \ \gg \ -t_1, -t_2 \ \sim \ 1\, . \ee
When written in light-cone coordinates, the asymptotical behavior
of our six momentum vectors $p_3 = (p_i^+,p_i^-,p^2_i,p_i^3)$ may
be parametrized by
\be \begin{array}{rclrcl}  p_1 &\sim& (-\omega^{-1},-\omega,
\ldots) \quad , & \quad
 p_2
&\sim& (-\omega, -\omega^{-1}, \ldots)\,,
\nonumber \\[2mm]
p_3 &\sim& (\omega, \omega^{-1}, \ldots) \quad  ,& \quad  p_4
&\sim& ( \omega^{ a},  \omega^{- a}, \ldots)\,,
\nonumber \\[2mm]
p_5 &\sim& ( \omega^{- b},  \omega^{ b}, \ldots) \quad , & \quad
p_6 &\sim& (\omega^{-1}, \omega, \ldots)\,, \end{array} \ee
where $\omega$ is a parameter that becomes large in the Regge
limit and we have only displayed the leading terms in the limit
$\omega \rightarrow \infty$. The two exponents $a$,$b$ are free
finite parameters. Note that all transverse momenta must be finite
to satisfy the on-shell condition. The dependence of kinematic
invariants on $\omega$ is now easy to compute
\be \begin{array}{rclrclrcl}  s  & \sim & \omega^2 \quad , & \quad
 s_{456} &  \sim & \omega^{(1+a)}\quad , &  \quad  s_{345} & \sim
 & \omega^{(1+b)}\,, \nonumber \\[2mm]
s_1 & \sim & \omega^{(1-a)} \quad , & \quad  s_2 &  \sim &
\omega^{(a+b)} \quad , & \quad  s_3 & \sim &  \omega^{(1-b)}\,,
\nonumber \\[2mm]
-t_1  & \sim  &   \omega^0 \quad , & \quad  -t_2 & \sim & \omega^0
\quad , & \quad  -s_{234} & \sim &  \omega^0\ . \end{array} \ee
Depending on our choice of $a$ and $b$ we can realize various
asymptotics of the kinematic invariants. If $a=b=1$, for example,
one finds $s_2 \sim s$. A regime with $s_1\sim s_2 \sim s_3$ is
reached in case we set $a=b=1/3$. Setting $a=b=0$, on the other
hand, allows us to perform a quasi multi-Regge limit in which
$s_2 \sim 1$.

From the scaling properties of kinematic invariants we can easily
deduce the following behavior of the cross-ratios
\eqref{eq:crossratio},
\be u_1 \ \sim \ \omega^{-(a+b)} \ \sim \ s_2^{-1} \quad , \quad
u_2 \ \sim\ \omega^{-(a+b)}\  \sim \  s_2^{-1}\ . \ee
Thereby we have shown that $u_1$ and $u_2$ vanish at the same
order when we take our Regge limit. Their ratio $u_1/u_2$ remains
constant. It remains to study the properties of $u_3$. Using the
Gram determinant relation to express $s$ in terms of the other
eight kinematic invariants we find
\be u_3 \ = \ \frac{s_2 s}{s_{456} s_{345}} = 1-f(
s_i,t_i,s_{i,i+1,i+2} )\ . \ee
After substitution of the gluon momenta we obtain
\be f(s_i,t_i,s_{i,i+1,i+2}) \ \sim\ \omega^{-(a+b)} \ \sim \
s_2^{-1}\,. \ee
Hence, both $u_1/(1-u_3)$ and $u_2/(1-u_3)$ remain finite for
$\omega \rightarrow \infty$. We have used these two ratios to
parametrize amplitudes in the Regge limit.

\section{Hirota equations}
\label{ap:he}

There exists another important set of functions $T_{s,m}$ that
appear in the context of gluon scattering. Our discussion in this
appendix will be kept general, i.e.\ it applies to an arbitrary
number $n$ of gluons. The functions $T_{s,m} = T_{s,m}(\theta)$
are parametrized by two integers $s,m$. They are subject to the
following set of Hirota equations,
\be T_{s,m}^+ T_{4-s,m}^-\ = \ T_{4-s,m+1}  T_{s,m-1}+ T_{s+1,m}
T_{s-1,m}\,, \label{Hirota} \ee
for $s=1,2,3$. The superscipts $\pm$ indicate shifts of the
spectral parameter, i.e.~
$$ f^{\pm}(\theta)\ =\ f^{[\pm 1]}(\theta) \ \ \mbox{ with }
\ \   f^{[m]}(\theta)\ =\ f(\theta + i m \ft{\pi}{4})\ \ . $$
We complete the description of the functions $T_{s,m}$ by a list
of boundary conditions.
\be T_{s,-1}\ =\ T_{s,n-4}=0 \,, \quad \mbox{where} \quad s=1,2,3
\ee and \be T_{s,0} \ =\ T_{0,m}\ =\ T_{4,m}\ =\ 1\ . \ee
Here, $n$ is the number of gluons. After imposing boundary
conditions, we are left with $3(n-5)$ functions.

The $T_{s,m}$ are related to $Y_{s,m}$ we have used throughout our
discussion through the relations
\be Y_{s,m}\ \equiv \ \frac{T_{s,m+1}
T_{4-s,m-1}}{T_{s+1,m}T_{s-1,m}} \ . \ee
Inserting these into the Hirota equations \eqref{Hirota} we
obtain
\be \frac{Y_{s,m}^- Y_{4-s,m}^+}{Y_{s+1,m}Y_{s-1,m}} \ = \ \frac{
(1+Y_{s,m+1})(1+Y_{4-s,m-1}) }{ (1+Y_{s+1,m})(1+Y_{s-1,m}) }\,,
\ee
where $s=1,2,3$ and $m=1,2,\ldots,n-5$. The behavior of $T_{s,m}$
on the boundary translate into the following boundary conditions
for $Y_{s,m}$,
\be Y_{0,m}\ =\ Y_{4,m}= \infty\,, \quad \mbox{and} \quad Y_{s,0}\
=\ Y_{s,n-4}=0 \quad \mbox{with} \quad s=1,2,3 \ . \ee
Using the set of $Y-$functions one may relate the parameters $m_m$,
$\phi_m$, $C_m$ of the $Y-$system through cross ratios. In order to
do so, one has to invert the following relations
\be \frac{ x_{p-k-1,p+k}^2 x_{p-k-2,p+k-1}^2 }{ x_{p-k-2,p+k}^2
x_{p-k-1,p+k-1}^2 } \
 =\ \left(\,  1+\frac{1}{Y_{2,2k-1}}\, \right)_{\theta=\frac{i \pi}{4}(2p-1)}\ = \
 \left( \, \frac{T_{2,2k-1}^+ T_{2,2k-1}^-}{T_{2,2k} T_{2,2k-3}}
 \, \right)_{\theta=\frac{i \pi}{4} (2p-1)}\,. \ee
Our equations \eqref{ufrompar} are recovered as a special case
when $n = 6$ with $k=1$ and $p = -1,0,1$. We have pointed out
before that the $Y$-functions satisfy the usual non-linear
equations as long as their argument $\theta$ remains inside the
strip $|\Im(\theta)|\leq \pi/4$. Whenever $\theta$ crosses one of
the lines $\Im(\theta)= i r \pi/4$, one has to add contributions
from poles of the kernel functions $K_i$. To avoid these
complications, we can also solve the original equations inside the
fundamental strip and extend the solutions to the complex plane
recursively, with the help of the $Y-$system equation, i.e.\
\be Y_{s,m}\ =\ \frac{(1+Y_{s,m+1}^{[1]})(1+Y_{4-s,m-1}^{[1]})}{
Y_{4-s,m}^{[2]}(1+1/Y_{s+1,m}^{[1]})(1+1/Y_{s-1,m}^{[1]})}
\label{eq:Hirota} \ . \ee
We have used this recursive construction of $Y_{s} = Y_{s,1}$ is
section 3.1 to obtain the formula \eqref{uYrelations2} for the
cross ratio $u_3$.

\section{Comments on numerical computations}

In this work, numerical calculations were used to find the path of
the analytical continuation in the space of parameters
$m,\phione,C$ and to check which solutions of $Y_i = -1$  cross
the integration contour. Most of this was done with the help of
Mathematica.

In order to determine the $Y-$functions for fixed $m$, $\phi$ and
$C$ we use an algorithm similar to one presented in
\cite{Alday:2010vh} for the case of $AdS_3$. It involves solving
the non-linear integral equations \eqref{NLIE1} and \eqref{NLIE2}
perturbatively for large enough $m$. The leading solution is given
by expressions that do not contain integrals, namely our eqs.\
\eqref{largemYtwo} and \eqref{largemYone}. To obtain the solution
with better precision we insert the $k^{th}$ iteration into the
integrals to obtain, i.e.\ schematically
\be \log Y_i^{(k+1)} (\theta)\ = \ -m_i \cosh(\theta - i \phi)+C_i
+ (K_{i j} \star\log(1+ Y_i^{(k)})) (\theta)\,. \ee
We perform $k$ such iterations until the $Y-$functions stabilize.
As a result we obtain $Y-$functions of real $\theta$, or rather a
discrete subset thereof. Moreover, using the integral equations we
can calculate the $Y-$functions for complex value of $\theta$ by
convoluting the kernel functions $K_i$ with the $Y-$functions of
real $\theta$. Due to existence of the poles in the kernels, we
must limit this evaluation of the $Y$-functions in the complex
$\theta$-plane to the strip $|\Im \theta| < \ft{\pi}{4}$. For
values of $\theta$ outside this strip, we recursively use Hirota
equations \eqref{eq:Hirota}. In this way, we can construct
$Y-$functions on the whole complex $\theta-$plane.

Next, we evaluate the $Y-$functions for special values of
$\theta=\pm \pi i/4,{3\pi i/4}$ which determine the cross ratios
$u_i$ defined in eq.\ \eqref{ufrompar}. Actually, for these values
of $\theta$ the numerical algorithms are not well convergent and
it is easier at these points to interpolate the $Y-$functions. But
at the end one can relate values of $m$, $\phi$ and $C$ to values
$u_i$. In order to find $Y-$functions along our analytical
continuations contour defined by eq.\ \eqref{pathu} one has to
invert the relation between the parameters $m$, $\phi$ and $C$ and
the cross ratios $u_i$. The values of $C$ along the path can be
found analytically, namely it is given by eq.\ \eqref{uthree}. The
other two parameters $m$ and $\phi$, on the other hand, require
numerical calculations.
An example of such path is shown in
figs.~\ref{fig:cpath},\ref{fig:cpath2}. Here, it has been
performed by solving a set of nonlinear equations \eqref{ufrompar}
using the Newton method.

Furthermore, one can also determine the positions of solutions to
the equations $Y_i =-1$ numerically, see figs. 4,5. These
solutions are moving while we are continuing along the contour in
parameter space. In the process of this continuation, some
solutions of $Y_3=-1$ and $Y_1 = -1$ cross the real line. This
happens roughly when $\varphi \sim \pi/2$. At this point we must
replace the original non-linear equations by those spelled out in
eqs.\ \eqref{eq:ane2twopoles} and \eqref{eq:ane13twopoles}.

The numerical results show that during analytical continuations
only from of the solutions to $Y_i = -1$ cross the integral
contour, see figs.\ 4 and 5. Moreover, to a very good
approximation, the imaginary parts of $m$ and $\phi$ behave
linearly with $\varphi$ while their real parts change
quadratically, compare also fig.~\ref{fig:cpath}. At the end of
the contour path, $m$ and $\phi$ go to values which are captured
by our eqs.\ \eqref{mprimem} and \eqref{phiprimephi}.

\bibliography{excitedhex}

\providecommand{\href}[2]{#2}\begingroup\raggedright\begin{thebibliography}{10}

\bibitem{Polyakov:1980ca}
A.~M. Polyakov, {\it {Gauge Fields as Rings of Glue}},  {\em Nucl. Phys.} {\bf
  B164} (1980) 171--188.

\bibitem{Maldacena:1997re}
J.~M. Maldacena, {\it {The large N limit of superconformal field theories and
  supergravity}},  {\em Adv. Theor. Math. Phys.} {\bf 2} (1998) 231--252,
  [\href{http://xxx.lanl.gov/abs/hep-th/9711200}{{\tt hep-th/9711200}}].

\bibitem{Witten:1998qj}
E.~Witten, {\it {Anti-de Sitter space and holography}},  {\em Adv. Theor. Math.
  Phys.} {\bf 2} (1998) 253--291,
  [\href{http://xxx.lanl.gov/abs/hep-th/9802150}{{\tt hep-th/9802150}}].

\bibitem{Gubser:1998bc}
S.~S. Gubser, I.~R. Klebanov, and A.~M. Polyakov, {\it {Gauge theory
  correlators from non-critical string theory}},  {\em Phys. Lett.} {\bf B428}
  (1998) 105--114, [\href{http://xxx.lanl.gov/abs/hep-th/9802109}{{\tt
  hep-th/9802109}}].

\bibitem{Arutyunov:2009zu}
G.~Arutyunov and S.~Frolov, {\it {String hypothesis for the $AdS_5 \times S^5$
  mirror}},  {\em JHEP} {\bf 03} (2009) 152,
  [\href{http://xxx.lanl.gov/abs/0901.1417}{{\tt arXiv:0901.1417}}].

\bibitem{Gromov:2009tv}
N.~Gromov, V.~Kazakov, and P.~Vieira, {\it {Exact Spectrum of Anomalous
  Dimensions of Planar N=4 Supersymmetric Yang-Mills Theory}},  {\em Phys. Rev.
  Lett.} {\bf 103} (2009) 131601,
  [\href{http://xxx.lanl.gov/abs/0901.3753}{{\tt arXiv:0901.3753}}].

\bibitem{Bombardelli:2009ns}
D.~Bombardelli, D.~Fioravanti, and R.~Tateo, {\it {Thermodynamic Bethe Ansatz
  for planar AdS/CFT: a proposal}},  {\em J. Phys.} {\bf A42} (2009) 375401,
  [\href{http://xxx.lanl.gov/abs/0902.3930}{{\tt arXiv:0902.3930}}].

\bibitem{Gromov:2009bc}
N.~Gromov, V.~Kazakov, A.~Kozak, and P.~Vieira, {\it {Exact Spectrum of
  Anomalous Dimensions of Planar N = 4 Supersymmetric Yang-Mills Theory: TBA
  and excited states}},  {\em Lett. Math. Phys.} {\bf 91} (2010) 265--287,
  [\href{http://xxx.lanl.gov/abs/0902.4458}{{\tt arXiv:0902.4458}}].

\bibitem{Arutyunov:2009ur}
G.~Arutyunov and S.~Frolov, {\it {Thermodynamic Bethe Ansatz for the $AdS_5
  \times S^5$ Mirror Model}},  {\em JHEP} {\bf 05} (2009) 068,
  [\href{http://xxx.lanl.gov/abs/0903.0141}{{\tt arXiv:0903.0141}}].

\bibitem{Hegedus:2009ky}
A.~Hegedus, {\it {Discrete Hirota dynamics for AdS/CFT}},  {\em Nucl. Phys.}
  {\bf B825} (2010) 341--365, [\href{http://xxx.lanl.gov/abs/0906.2546}{{\tt
  arXiv:0906.2546}}].

\bibitem{Gromov:2009zb}
N.~Gromov, V.~Kazakov, and P.~Vieira, {\it {Exact Spectrum of Planar ${\cal
  N}=4$ Supersymmetric Yang- Mills Theory: Konishi Dimension at Any Coupling}},
   {\em Phys. Rev. Lett.} {\bf 104} (2010) 211601,
  [\href{http://xxx.lanl.gov/abs/0906.4240}{{\tt arXiv:0906.4240}}].

\bibitem{Arutyunov:2009ux}
G.~Arutyunov and S.~Frolov, {\it {Simplified TBA equations of the $AdS_5 \times
  S^5$ mirror model}},  {\em JHEP} {\bf 11} (2009) 019,
  [\href{http://xxx.lanl.gov/abs/0907.2647}{{\tt arXiv:0907.2647}}].

\bibitem{Gromov:2009tq}
N.~Gromov, {\it {Y-system and Quasi-Classical Strings}},  {\em JHEP} {\bf 01}
  (2010) 112, [\href{http://xxx.lanl.gov/abs/0910.3608}{{\tt
  arXiv:0910.3608}}].









\bibitem{Bern:2005iz}
Z.~Bern, L.~J. Dixon, and V.~A. Smirnov, {\it {Iteration of planar amplitudes
  in maximally supersymmetric Yang-Mills theory at three loops and beyond}},
  {\em Phys. Rev.} {\bf D72} (2005) 085001,
  [\href{http://xxx.lanl.gov/abs/hep-th/0505205}{{\tt hep-th/0505205}}].

\bibitem{Drummond:2007au}
J.~M. Drummond, J.~Henn, G.~P. Korchemsky, and E.~Sokatchev, {\it {Conformal
  Ward identities for Wilson loops and a test of the duality with gluon
  amplitudes}},  {\em Nucl. Phys.} {\bf B826} (2010) 337--364,
  [\href{http://xxx.lanl.gov/abs/0712.1223}{{\tt arXiv:0712.1223}}].

\bibitem{Drummond:2007bm}
J.~M. Drummond, J.~Henn, G.~P. Korchemsky, and E.~Sokatchev, {\it {The hexagon
  Wilson loop and the BDS ansatz for the six- gluon amplitude}},  {\em Phys.
  Lett.} {\bf B662} (2008) 456--460,
  [\href{http://xxx.lanl.gov/abs/0712.4138}{{\tt arXiv:0712.4138}}].

\bibitem{Bern:2008ap}
Z.~Bern {\em et.~al.}, {\it {The Two-Loop Six-Gluon MHV Amplitude in Maximally
  Supersymmetric Yang-Mills Theory}},  {\em Phys. Rev.} {\bf D78} (2008)
  045007, [\href{http://xxx.lanl.gov/abs/0803.1465}{{\tt arXiv:0803.1465}}].

\bibitem{Bartels:2008ce}
J.~Bartels, L.~N. Lipatov, and A.~Sabio~Vera, {\it {BFKL Pomeron, Reggeized
  gluons and Bern-Dixon-Smirnov amplitudes}},  {\em Phys. Rev.} {\bf D80}
  (2009) 045002, [\href{http://xxx.lanl.gov/abs/0802.2065}{{\tt
  arXiv:0802.2065}}].

\bibitem{Bartels:2008sc}
J.~Bartels, L.~N. Lipatov, and A.~Sabio~Vera, {\it {N=4 supersymmetric Yang
  Mills scattering amplitudes at high energies: the Regge cut contribution}},
  {\em Eur. Phys. J.} {\bf C65} (2010) 587--605,
  [\href{http://xxx.lanl.gov/abs/0807.0894}{{\tt arXiv:0807.0894}}].

\bibitem{Schabinger:2009bb}
R.~M. Schabinger, {\it {The Imaginary Part of the N = 4 Super-Yang-Mills
  Two-Loop Six-Point MHV Amplitude in Multi-Regge Kinematics}},  {\em JHEP}
  {\bf 11} (2009) 108, [\href{http://xxx.lanl.gov/abs/0910.3933}{{\tt
  arXiv:0910.3933}}].

\bibitem{Lipatov:2010qg}
L.~N. Lipatov and A.~Prygarin, {\it {Mandelstam cuts and light-like Wilson
  loops in N=4 SUSY}},  \href{http://xxx.lanl.gov/abs/1008.1016}{{\tt
  arXiv:1008.1016}}.

\bibitem{Goncharov:2010jf}
A.~B. Goncharov, M.~Spradlin, C.~Vergu, and A.~Volovich, {\it {Classical
  Polylogarithms for Amplitudes and Wilson Loops}},
  \href{http://xxx.lanl.gov/abs/1006.5703}{{\tt arXiv:1006.5703}}.

\bibitem{DelDuca:2009au}
V.~Del~Duca, C.~Duhr, and V.~A. Smirnov, {\it {An Analytic Result for the
  Two-Loop Hexagon Wilson Loop in N = 4 SYM}},  {\em JHEP} {\bf 03} (2010) 099,
  [\href{http://xxx.lanl.gov/abs/0911.5332}{{\tt arXiv:0911.5332}}].

\bibitem{DelDuca:2010zg}
V.~Del~Duca, C.~Duhr, and V.~A. Smirnov, {\it {The Two-Loop Hexagon Wilson Loop
  in N = 4 SYM}},  {\em JHEP} {\bf 05} (2010) 084,
  [\href{http://xxx.lanl.gov/abs/1003.1702}{{\tt arXiv:1003.1702}}].

\bibitem{Lipatov:2009nt}
L.~N. Lipatov, {\it {Integrability of scattering amplitudes in N=4 SUSY}},
  {\em J. Phys.} {\bf A42} (2009) 304020,
  [\href{http://xxx.lanl.gov/abs/0902.1444}{{\tt arXiv:0902.1444}}].

\bibitem{Alday:2007hr}
L.~F. Alday and J.~M. Maldacena, {\it {Gluon scattering amplitudes at strong
  coupling}},  {\em JHEP} {\bf 06} (2007) 064,
  [\href{http://xxx.lanl.gov/abs/0705.0303}{{\tt arXiv:0705.0303}}].

\bibitem{Alday:2009yn}
L.~F. Alday and J.~Maldacena, {\it {Null polygonal Wilson loops and minimal
  surfaces in Anti- de-Sitter space}},  {\em JHEP} {\bf 11} (2009) 082,
  [\href{http://xxx.lanl.gov/abs/0904.0663}{{\tt arXiv:0904.0663}}].

\bibitem{Alday:2009dv}
L.~F. Alday, D.~Gaiotto, and J.~Maldacena, {\it {Thermodynamic Bubble Ansatz}},
   \href{http://xxx.lanl.gov/abs/0911.4708}{{\tt arXiv:0911.4708}}.

\bibitem{Alday:2010vh}
L.~F. Alday, J.~Maldacena, A.~Sever, and P.~Vieira, {\it {Y-system for
  Scattering Amplitudes}},  \href{http://xxx.lanl.gov/abs/1002.2459}{{\tt
  arXiv:1002.2459}}.

\bibitem{Dorey:1996re}
P.~Dorey and R.~Tateo, {\it {Excited states by analytic continuation of TBA
  equations}},  {\em Nucl. Phys.} {\bf B482} (1996) 639--659,
  [\href{http://xxx.lanl.gov/abs/hep-th/9607167}{{\tt hep-th/9607167}}].

\bibitem{Dorey:1997rb}
P.~Dorey and R.~Tateo, {\it {Excited states in some simple perturbed conformal
  field theories}},  {\em Nucl. Phys.} {\bf B515} (1998) 575--623,
  [\href{http://xxx.lanl.gov/abs/hep-th/9706140}{{\tt hep-th/9706140}}].

\bibitem{Alday:2010ku}
L.~F. Alday, D.~Gaiotto, J.~Maldacena, A.~Sever, and P.~Vieira, {\it {An
  Operator Product Expansion for Polygonal null Wilson Loops}},
  \href{http://xxx.lanl.gov/abs/1006.2788}{{\tt arXiv:1006.2788}}.

\bibitem{Hatsuda:2010vr}
Y.~Hatsuda, K.~Ito, K.~Sakai, and Y.~Satoh, {\it {Six-point gluon scattering
  amplitudes from $Z_4$-symmetric integrable model}},  {\em JHEP} {\bf 09}
  (2010) 064, [\href{http://xxx.lanl.gov/abs/1005.4487}{{\tt
  arXiv:1005.4487}}].

\bibitem{GrRy}
I.~Gradshteyn and I.~Ryzhik, {\em {Table of integrals, series, and products.
  Translated from the Russian. Translation edited and with a preface by Alan
  Jeffrey and Daniel Zwillinger. 6th ed.}}
\newblock {San Diego, CA: Academic Press. xlvii, 1163 p.}, 2000.

\end{thebibliography}\endgroup

\end{document}